# Hypercube quantum search: exact computation of the probability of success in polynomial time

Hugo Pillin[1,2] · Gilles Burel[1] · Paul Baird[2] · El-Houssain Baghious[1] · Roland Gautier[1]

## Abstract

In the emerging domain of quantum algorithms, Grover's quantum search is certainly one of the most significant. It is relatively simple, performs a useful task and more importantly, does it in an optimal way. However, due to the success of quantum walks in the field, it is logical to study quantum search variants over several kinds of walks. In this paper, we propose an in-depth study of the quantum search over a hypercube layout. First, through the analysis of elementary walk operators restricted to suitable eigenspaces, we show that the acting component of the search algorithm takes place in a small subspace of the Hilbert workspace that grows linearly with the problem size. Subsequently, we exploit this property to predict the exact evolution of the probability of success of the quantum search in polynomial time.

Gilles Burel, Paul Baird, El-Houssain Baghious and Roland Gautier have contributed equally to this work.

✉ Hugo Pillin
  Hugo.Pillin@univ-brest.fr

  Gilles Burel
  Gilles.Burel@univ-brest.fr

  Paul Baird
  Paul.Baird@univ-brest.fr

  El-Houssain Baghious
  El-Houssain.Baghious@univ-brest.fr

  Roland Gautier
  Roland.Gautier@univ-brest.fr

1   Lab-STICC, University of Brest, 6 Avenue Le Gorgeu, 29200 Brest, France

2   LMBA, University of Brest, 6 Avenue Le Gorgeu, 29200 Brest, France

                                                                                                                             Springer

## Introduction

Many computational problems can be reduced to the search of one or more items in a set that meet a predefined criterion. For instance in a digital communications context, with channel coding using block codes, the receiver has to find the binary word which best explains the received data. If there are 50 bits per data word, this means finding the best solution among $2^{50}$ (approximately $10^{15}$) possibilities.

This is the kind of problem Grover's algorithm, introduced in [4], can resolve with a high probability in $\mathcal{O}(\sqrt{N})$ steps, $N$ being the number of possibilities. This implies that quantum computing could significantly speed up some problems too complex to be simulated with a classical computer.

Quantum walks over graphs are a great tool in the design of quantum algorithms. It has been shown in [1] that quantum walks can be faster than their classical counterparts up to a polynomial factor. Further reading about quantum walks can be found in [9].

The choice of the hypercube structure over other walks is motivated by its direct connection with information science, where it is often used to represent binary values. Indeed, every $n$-bit value can be placed on the $n$-dimensional hypercube, where the number of edges separating the two values is equal to the number of bits that differ from one value to the other. Additionally, a displacement along the $d$-th dimension on the hypercube is equivalent to a flip (0 becomes 1 and conversely) of the value of the $d$-th bit, as illustrated in Fig. 1. It has been shown in [5] that quantum walks on hypercube are faster to mix than a classical random walk.

While Grover's algorithm is simple to build, one is also required to determine the optimal number of iterations to get the maximal probability of success, since if there are too many iterations, the probability decreases. The optimal number of iterations for the standard Grover's algorithm $R$ is bounded by:

$$R \leq \left\lceil \frac{\pi}{4} \sqrt{\frac{N}{M}} \right\rceil, \tag{0.1}$$

where $M$ is the number of solutions. Note that even if the number of solutions is unknown, it is possible to use an incremental procedure as shown in [8] or to compute an approximation of this number using the quantum counting technique. However, this formula is invalid for its quantum walk counterparts. Asymptotic results have been obtained by Shenvi et al. in [7] for a search by quantum walks on the hypercube for which there is a unique solution. They prove that when the hypercube dimension $n$ becomes large, after $\frac{\pi}{2}\sqrt{2^n - 1}$ iterations, the probability of success is $\frac{1}{2} - \mathcal{O}(\frac{1}{n})$. In this paper, we describe a method that allows us to compute the exact probability of success of a walk on the hypercube for any number of solutions, in order to find the optimal number of iterations required to maximize this probability.

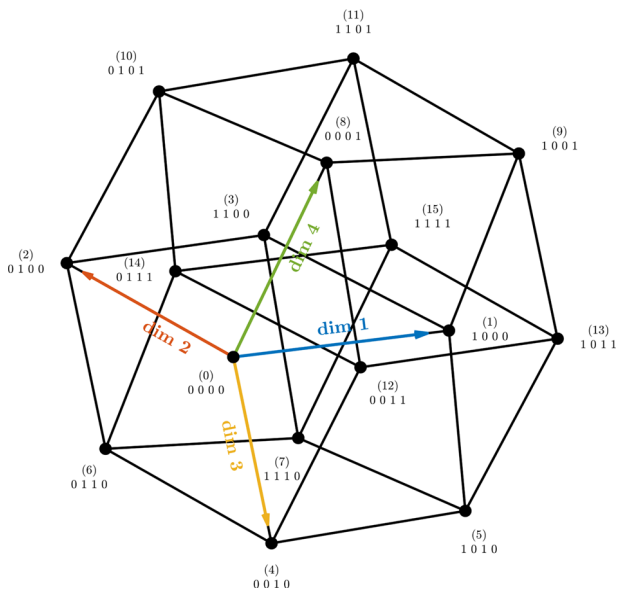

**Fig. 1** The 4-dimensional hypercube structure, with vertices labeled and direction arrowed

The main idea behind the quantum walks is to split the Hilbert space $\mathcal{H}$ into two distinct parts: the position space $\mathcal{H}^\mathcal{S}$, which corresponds to the vertices on the graph (in our case, a hypercube), and the coin space $\mathcal{H}^\mathcal{C}$, which corresponds to the possible directions of displacement from the vertices. Thus, $\mathcal{H} = \mathcal{H}^\mathcal{S} \otimes \mathcal{H}^\mathcal{C}$, where $\otimes$ is the Kronecker tensor product. On an $n$-dimensional hypercube, there are $N = 2^n$ vertices and $n$ possible directions. In this representation, we chose as basis states the distinct positions in $\mathcal{H}^\mathcal{S}$ and the distinct directions in $\mathcal{H}^\mathcal{C}$. Therefore, a basis state in $\mathcal{H}$ can be written $|\psi\rangle = |p\rangle|d\rangle$, where $|p\rangle$ is the position state and $|d\rangle$ the direction of movement, often referred as the "quantum coin". In the $n$-dimension hypercube, $p$ and $d$ are integers with $p \in [0, N-1]$ and $d \in [1, n]$. On a hypercube, it is often easier to use binary words to index position. For example, the state located on vertex 4 will be displaced along direction 1 that is $|\psi\rangle = |4\rangle|1\rangle$. For each value of $p$, we associate a binary word $\rho$. Our last example can be written as $|\psi\rangle = |0010\rangle|1\rangle$. Note that the most significant bit of $\rho$ is on the right and that moving along the $i$-th dimension will therefore flip the $i$-th bit of $\rho$.

The discrete-time quantum walk search algorithm consists in the repeated application of two unitary operators, the oracle $O$ followed by the uniform walk operator $U$. The global iteration operator is $Q = UO$. The oracle's role is to mark the solutions, while walk operator disperses the states on the hypercube. The uniform walk operator $U$ is itself the product of two operators: the shift operator $S$ and the coin operator $C$, so that $U = SC$ and $Q = SCO$. These operators are defined and studied in Sect. 2.

Let $|\psi_0\rangle$ be the initial state and $t$ the number of iterations until the system is measured. At each iteration, the operator $Q$ is applied. This operator is constant along the process and is defined once according to the problem. Therefore, the final state will be:

$$|\psi_t\rangle = Q^t |\psi_0\rangle. \tag{0.2}$$

The algorithm is a success if the measurement of $|\psi_t\rangle$ returns one of the $M$ solutions. As for all quantum algorithms, the outcome is probabilistic. Let $|s\rangle$ be the uniform superposition of all solutions. The probability of success after $t$ iterations is:

$$p_t = |\langle s|\psi_t\rangle|^2. \tag{0.3}$$

Note that this expression, as well as many other fundamental results about quantum information, can be found in [6].

The goal of this paper is to maximize $p_t$. As $Q$ depends only on the size of the problem and its solutions, the probability of success depends only on $t$. In order to maximize $p_t$, we study a special subspace of $\mathcal{H}$, which is the complement of the joint eigenspace of the operators $U$ and $O$. Let us denote this joint eigenspace $\overline{\mathcal{E}}$ and its complement $\mathcal{E}$. Let $U_{\overline{\mathcal{E}}}$, $O_{\overline{\mathcal{E}}}$ and $Q_{\overline{\mathcal{E}}}$ be the components of $U$, $O$ and $Q$ inside $\overline{\mathcal{E}}$. In their joint eigenspace, operators always commute, and as we will see in Sect. 2.3, $O$ is always self-inverse. Therefore, we have

$$Q_{\overline{\mathcal{E}}}^2 = \left(U_{\overline{\mathcal{E}}} O_{\overline{\mathcal{E}}}\right)\left(U_{\overline{\mathcal{E}}} O_{\overline{\mathcal{E}}}\right), \tag{0.4}$$

$$= U_{\overline{\mathcal{E}}} O_{\overline{\mathcal{E}}}^2 U_{\overline{\mathcal{E}}}, \tag{0.5}$$

$$= U_{\overline{\mathcal{E}}}^2. \tag{0.6}$$

This means that in $\overline{\mathcal{E}}$, to apply the global operator twice is to apply the uniform walk twice, without the search component brought by the oracle. Therefore, it has no reason to converge to a solution, and the algorithm effectively takes place in $\mathcal{E}$.

We proceed as follows:

- We show that the dimension of $\mathcal{E}$ is approximately $2Mn$, and its exact value will be given in Sect. 6. That means that $\mathcal{E}$ grows linearly in $n$, while $\mathcal{H}$ grows exponentially. We also prove that the uniform state $|u\rangle$ and the superposition of solutions $|s\rangle$ are both in $\mathcal{E}$.
- We determine the eigenvalues associated with $Q$ in $\mathcal{E}$ and their multiplicities. The number of eigenvalues to determine is lesser than or equal to $2Mn$. We also have developed a method which can be easily programmed and executed on a classical computer, which uses a minima search over a specifically designed criterion.
- Finally, we propose a method to compute the components in $\mathcal{E}$ of the initial state and of the uniform superposition of the solutions $|s\rangle$. With these components and the eigenvalues, it is then possible to compute the probability of success with respect to the number of iterations in polynomial time.

# 1 Notations and mathematical background

## 1.1 Notations

Throughout this paper, some notations will frequently be used in different contexts:

- $n$ is the dimension of the hypercube and the dimension of the coin space $\mathcal{H}^C$.
- $N = 2^n$ is the number of positions on the hypercube and the dimension of the position space $\mathcal{H}^S$.

- $N_e = nN$ is the dimension of the global space $\mathcal{H} = \mathcal{H}^{\mathcal{S}} \otimes \mathcal{H}^{\mathcal{C}}$.
- $M$ is the number of solutions for a given problem.
- $A^{\mathrm{T}}$ is the transpose of the matrix $A$ and $A^{\dagger}$ is the conjugate transpose of the matrix $A$.
- $|u_k\rangle$ is the $k \times 1$ uniform column vector. We have

$$|u_k\rangle = \frac{1}{\sqrt{k}} \begin{bmatrix} 1 \\ 1 \\ \vdots \\ 1 \end{bmatrix}. \tag{1.1}$$

When there is no ambiguity, we may omit the index and simply write $|u\rangle$. We also define the $n \times (n-1)$ $\Lambda$ matrix, characterized by $\langle u|\Lambda = 0$ and $\Lambda^{\mathrm{T}} \Lambda = 0$.
- $|1_p\rangle$ is a length $N$ vector which contains 1 at the position $p$ and 0 elsewhere.
- $I$ is the $2 \times 2$ identity matrix, while $I_n$ is the $n \times n$ identity matrix.
- $X$ is the Pauli-$X$ matrix, also known as the quantum NOT:

$$X = \begin{bmatrix} 0 & 1 \\ 1 & 0 \end{bmatrix}. \tag{1.2}$$

- $H$ is the normalized Hadamard matrix:

$$H = \frac{1}{\sqrt{2}} \begin{bmatrix} 1 & 1 \\ 1 & -1 \end{bmatrix}. \tag{1.3}$$

We denote the tensor power $H^{\otimes n}$ by $H_N$, as it is frequently used.
We will also use the unnormalized Hadamard $\bar{H}_N$, given by

$$\bar{H}_N = \sqrt{N} H_N. \tag{1.4}$$

- $E_{\alpha}^{A}$ is the eigenspace of the operator $A$ associated with the eigenvalue $\alpha$. We denote the intersection of two eigenspaces $E_{\alpha}^{A} \cap E_{\beta}^{B}$ by $E_{\alpha,\beta}^{A,B}$. As almost every eigenvalue we will encounter is $\pm 1$, we will simply denote them by their signs, such as in $E_{+,-}^{A,B}$.
- $P_{a,b}$ is a permutation matrix obtained by reading column per column elements organized row per row. For example, if $a = 2$ and $b = 3$, we write all integers from 1 to $a \times b = 6$ row per row in a $a \times b = 2 \times 3$ matrix. We obtain:

$$\begin{bmatrix} 1 & 2 & 3 \\ 4 & 5 & 6 \end{bmatrix}. \tag{1.5}$$

By reading column per column, we get the order 1, 4, 2, 5, 3, 6. The permutation-associated matrix $P_{2,3}$ is obtained by moving the rows of a $6 \times 6$ identity matrix according to this order. The first row remains the first, the second row is moved to the fourth, etc. We obtain

$$P_{2,3} = \begin{bmatrix} 1 & & & & & \\ & 1 & & & & \\ & & & 1 & & \\ & & 1 & & & \\ & & & & 1 & \\ & & & & & 1 \end{bmatrix}. \tag{1.6}$$

- The default space we are using is $\mathcal{H} = \mathcal{H}^S \otimes \mathcal{H}^C$, but we will also need the $\mathcal{H}^C \otimes \mathcal{H}^S$ space. If $A$ is an $N_e \times N_e$ operator in $\mathcal{H}^S \otimes \mathcal{H}^C$, we denote its $\mathcal{H}^C \otimes \mathcal{H}^S$ counterpart by $A^{CS}$.

### 1.2 The Kronecker product

The Kronecker product is a bilinear operation on two matrices. Let $A$ and $B$ be two matrices of dimension $m_A \times n_A$ and $m_B \times n_B$. Let $a_{jk}$ be the $A$ element of the $j$-th row and $k$-th column. The tensor product of $A$ and $B$ is:

$$A \otimes B = \begin{bmatrix} a_{11}B & \cdots & a_{1n_A}B \\ \vdots & \ddots & \vdots \\ a_{m_A1}B & \cdots & a_{m_An_A}B \end{bmatrix}. \tag{1.7}$$

We will also use the Kronecker matrix power, the Kronecker product of a matrix with itself a given number of times. The $k$-th Kronecker power of a matrix $A$ is denoted $A^{\otimes k}$.

The inverse of a tensor product $A \otimes B$ is:

$$(A \otimes B)^{-1} = A^{-1} \otimes B^{-1}. \tag{1.8}$$

The interactions between the matrix product and the Kronecker product follow the property known as mixed product:

$$(A \otimes B)(C \otimes D) = (AC) \otimes (BD). \tag{1.9}$$

The Kronecker product is associative but not commutative. However, the structure of the permutation matrices defined in Sect. 1.1, inspired from [2], leads to this interesting property:

$$A \otimes B = P_{a,b}(B \otimes A) P_{b,a}. \tag{1.10}$$

Note that $P_{a,b}^T = P_{b,a}$, so

$$A \otimes B = P_{a,b}(B \otimes A) P_{a,b}^T. \tag{1.11}$$

The permutation matrices allow us to switch between the workspaces $\mathcal{H}^S \otimes \mathcal{H}^C$ and $\mathcal{H}^C \otimes \mathcal{H}^S$.

### 1.3 The discrete Fourier transform

One important use of the discrete Fourier transform (DFT) is to diagonalize certain operators. In its general form, the coefficients of the $N \times N$ DFT operator are given by:

$$\text{DFT}_{j,k} = \frac{\omega^{jk}}{\sqrt{N}} \text{ with } j, k \in [0, N-1], \tag{1.12}$$

where $\omega = e^{-2\pi i/N}$ and $\text{DFT}_{j,k}$ is the value on the $j$-th row and $k$-th column.

In our case, we will use the DFT in $\mathcal{H}^S$, where each component belongs to $\{0, 1\}$. When $N = 2$, the DFT matrix is equal to the Hadamard matrix $H$. Therefore, applying the DFT over each of the $n$ dimensions of the hypercube in $\mathcal{H}^S$ is equivalent to applying $H_N$ to the state vector. We also define $F$ to be the $N_e \times N_e$ matrix that applies the DFT in $\mathcal{H}^S$ while leaving $\mathcal{H}^C$ unchanged:

$$F = H_N \otimes I_n. \tag{1.13}$$

In this paper, we will say that we switch between the Fourier and the original domains whenever we multiply by $F$. Note that since $H$ is self-inverse, $F$ is too, and we will therefore invert a Fourier transform with another one. Note that the use of the term "Fourier transform" in this paper implies different calculations: in $\mathcal{H}$, it is a multiplication by $F$, while in $\mathcal{H}^S$ it is the transform produced by $H_N$. Furthermore, the Fourier transform of an operator $A$ in $\mathcal{H}$ is $FAF$, while the one of a vector $|v\rangle$ (or a matrix which is a collection of vectors) if $F|v\rangle$. The same goes for $H_N$ in $\mathcal{H}^S$.

We will denote the Fourier transform of a matrix $A$ by $\hat{A}$.

### 1.4 The singular value decomposition

The singular value decomposition, or SVD, is a matrix factorization that generalizes the eigendecomposition. The SVD of a given complex $M \times N$ matrix $A$ is of the form

$$A = USV^\dagger, \tag{1.14}$$

where $U$ is a $M \times M$ unitary matrix, $S$ a $M \times N$ rectangular diagonal matrix and $V$ an $N \times N$ unitary matrix. The diagonal elements of $S$ are non-negative real values known as the singular values.

The SVD can always be computed in polynomial time. According to Golub and Van Loan in [3], the complexity of the R-SVD algorithm is $\mathcal{O}(N^3 + NM^2)$. It is even possible to compute only the singular values in $\mathcal{O}(MN^2)$. We will not discuss further the complexity of the SVD as it can already be considered as efficient.

In this paper, we only use a reduced version of this decomposition, namely the thin SVD. Let us define $K = \min(M, N)$. In this variant of the SVD, we keep only the first $K$ columns of $U$ and $S$ and the first $K$ rows of $S$ and $V$. This allows for a significant economy in computation time and memory, while the $A = USV^\dagger$ relation remains true.

### 1.5 Notes on unitary matrix

Unitary matrices are omnipresent in the quantum domain. As a reminder, an $n \times n$ matrix $A$ is said to be "unitary" when

$$A^\dagger A = A A^\dagger = I_n. \tag{1.15}$$

If a matrix is unitary, all its eigenvalues lie on the unit circle and all its eigenspaces are orthogonal. Those eigenvalues and eigenvectors are either real or come in conjugate pairs of complex numbers.

If $A$ is a unitary matrix such as $A^2 = I_n$, then all its eigenvalues are $\pm 1$. The dimensions of its eigenspaces are:

$$\dim E^A_\pm = \frac{n \pm \mathrm{tr}(A)}{2}, \tag{1.16}$$

where $E^A_+$ is the eigenspace associated with $+1$ and $E^A_-$ to $-1$. Furthermore, if $A^2 = I_n$, we can express the projector onto $E^A_\pm$ as

$$P^A_\pm = \frac{1}{2}\left(I_n \pm A\right). \tag{1.17}$$

### 1.6 Indexation and signatures

In this paper, we will often use the binary word associated with an index. We always use a Latin character to denote the index and a Greek one to denote its binary equivalent. For instance, if $p = 3$, then $\rho = 110$. Sometimes, we will use those binary indices as vectors. In such cases, we will use a ket to denote them, such as $|\rho\rangle$.

In order to simplify several proofs, we will also label elements (rows, columns, diagonal values) with the Hamming weight $w$ of their binary indices. For example, if $N = 8$, the indices are $000, 100, 010, 110, 001, 101, 011, 111$, and their respective weights are $0, 1, 1, 2, 1, 2, 2, 3$.

When working with $N_e \times N_e$ matrices, we will attribute a $\pm 1$ signature $\sigma$ to each row or column. We split the $N_e$ elements in $N$ intervals of size $n$. The $n$ signatures in the $p$-th interval are defined from the $n$-bit binary representation of $p$, where bits 0 are mapped to $\sigma = +1$, and 1 to $\sigma = -1$. For example, with $n = 3$, an $N_e \times N_e$ matrix has $N = 8$ blocks of 3 elements. In the case of the first block, $p = 0$ and its binary representation is $000$. Therefore, the three first rows or columns have a signature $\sigma = +1$. In the second block, the position binary representation is $100$ and the fourth element has a signature $\sigma = -1$, while the fifth and the sixth have a signature $\sigma = +1$. This apparently arbitrary definition will naturally emerge in Sect. 2.1.

### 1.7 Useful submatrices

Throughout this paper, we will use submatrices of $H_N$:

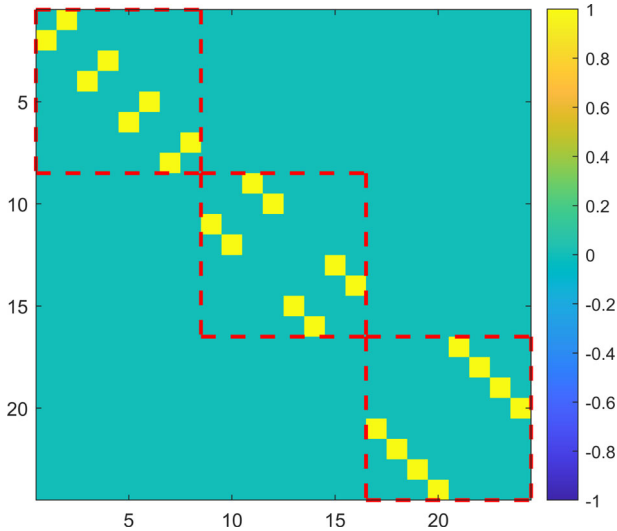

**Fig. 2** The shift operator $S^{CS}$ in $\mathcal{H}^C \otimes \mathcal{H}^S$ for $n = 3$. The $S_d$ blocks are delimited by the dashed lines

- $H_N^s$ is the $N \times M$ matrix obtained by keeping only the column of $H_N$ associated with the solutions.
- $H_N^{s,w}$ is the $\binom{n}{w} \times M$ matrix obtained by keeping only the rows of $H_N^s$ whose indices have a Hamming weight of $w$.
- We define $I_N^s$ and $I_N^{s,w}$ in the same manner from the $N \times N$ identity matrix.
- We define the $N \times (N - M)$ matrix $I_N^{\bar{s}}$ in a similar way to $I_N^s$, but by keeping only the columns of $I_N$ associated to the non-solutions.
- $\bar{H}_N^w$ is the $\binom{n}{w} \times N$ matrix obtained by keeping only the rows of $\bar{H}_N$ whose indices have a Hamming weight of $w$

## 2 Operators and their eigenspaces

### 2.1 The shift operator

The shift operator $S$ represents a position shift controlled by the quantum coin. Along the $d$-th dimension, the shift operator permutes the values associated with adjacent vertices. It follows that along direction $d$ it is simply the operator $X$. For each of the $n$ dimensions, we can create an $N \times N$ operator $S_d$ that only affects the $d$-th dimension in $\mathcal{H}^S$:

$$S_d = I^{\otimes(n-d)} \otimes X \otimes I^{\otimes(d-1)}. \quad (2.1)$$

From this, we can find the shift operator in the $\mathcal{H}^C \otimes \mathcal{H}^S$: it is a block diagonal matrix we denote $S^{CS}$, where each of its blocks is a $S_d$. We can identify the $n$ $S_d$ blocks in Fig. 2.

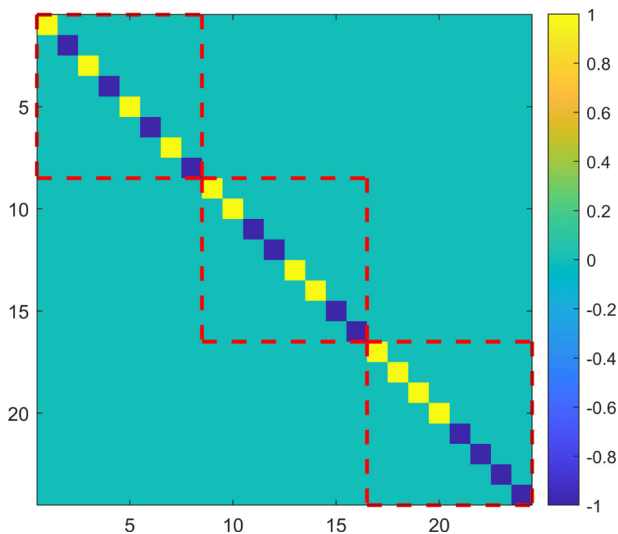

**Fig. 3** The diagonalized shift operator $\hat{S}^{\mathcal{CS}}$ in $\mathcal{H}^{\mathcal{C}} \otimes \mathcal{H}^{\mathcal{S}}$ for $n = 3$. The $\hat{S}_d$ blocks are delimited by the dashed lines

We can show that the $S_d$ are diagonalized by $H_N$. If we define $\hat{S}_d = H_N S_d H_N$, we have

$$\hat{S}_d = H_N S_d H_N, \tag{2.2}$$

$$= H^{\otimes n} \left( I^{\otimes (n-d)} \otimes X \otimes I^{\otimes (d-1)} \right) H^{\otimes n}, \tag{2.3}$$

$$= \left( H^2 \right)^{\otimes (n-d)} \otimes (HXH) \otimes \left( H^2 \right)^{\otimes (d-1)}. \tag{2.4}$$

$\hat{S}_d$ is therefore diagonal because $H^2 = I$ and $HXH$ is diagonal ($HXH$ is the Pauli Z matrix):

$$HXH = \begin{bmatrix} 1 & 0 \\ 0 & -1 \end{bmatrix}. \tag{2.5}$$

As a consequence, the diagonalized operator in $\mathcal{H}^{\mathcal{C}} \otimes \mathcal{H}^{\mathcal{S}}$, denoted $\hat{S}^{\mathcal{CS}}$, is composed of $n$ diagonal submatrices $\hat{S}_d$. We note that each $\hat{S}_d$ matrix is made up of $2^{n-d}$ repetitions of a block made of $2^{d-1}$ '+1' followed by $2^{d-1}$ '-1', as seen in Fig. 3. The relation between the shift operator and its diagonalized version in $\mathcal{H}^{\mathcal{C}} \otimes \mathcal{H}^{\mathcal{S}}$ is

$$\hat{S}^{\mathcal{CS}} = (I_n \otimes H_N) S^{\mathcal{CS}} (I_n \otimes H_N). \tag{2.6}$$

The shift operator has a simple block diagonal structure in $\mathcal{H}^{\mathcal{C}} \otimes \mathcal{H}^{\mathcal{S}}$, but we need to get its representation in $\mathcal{H}^{\mathcal{S}} \otimes \mathcal{H}^{\mathcal{C}}$, that we denote $S$. We will use the permutation matrices seen in Sect. 1.2 to obtain $S$. Let us note $P = P_{N,n}$. We have

$$S = P S^{\mathcal{CS}} P^T. \tag{2.7}$$

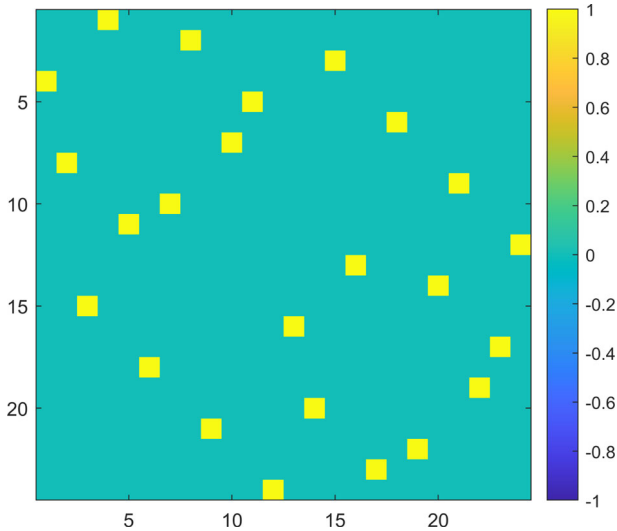

**Fig. 4** The shift operator $S$ in $\mathcal{H}^S \otimes \mathcal{H}^C$ for $n = 3$

As we can see in Fig. 4, the shift operator is more complex in $\mathcal{H}^S \otimes \mathcal{H}^C$. However, it is the opposite for the coin operator, which is block diagonal in $\mathcal{H}^S \otimes \mathcal{H}^C$ but not in $\mathcal{H}^C \otimes \mathcal{H}^S$. We arbitrarily choose to study the operators in $\mathcal{H}^S \otimes \mathcal{H}^C$, as this space is the most commonly used in quantum walk studies.

In order to study the eigendecomposition of $S$, we need to diagonalize it. Note that $P^T P = I_{N_e}$. We have

$$S = P S^{CS} P^T, \tag{2.8}$$
$$= P (I_n \otimes H_N) \hat{S}^{CS} (I_n \otimes H_N) P^T, \tag{2.9}$$
$$= P (I_n \otimes H_N) P^T P \hat{S}^{CS} P^T P (I_n \otimes H_N) P^T, \tag{2.10}$$
$$= (H_N \otimes I_n) \left( P \hat{S}^{CS} P^T \right) (H_N \otimes I_n), \tag{2.11}$$
$$= F \left( P \hat{S}^{CS} P^T \right) F. \tag{2.12}$$

As $\hat{S}^{CS}$ is diagonal, $P \hat{S}^{CS} P^T$ is diagonal too (that is true for any permutation matrix). Therefore, $F$ diagonalizes $S$ and we can denote $P \hat{S}^{CS} P^T$ by $\hat{S}$.

The structure of $\hat{S}$ is remarkable: if we map the $+1$ to 0 and the $-1$ to 1, we find along its diagonal all the possible $n$-bit words in ascending order, as seen in Fig. 5. We also observe that those values are equivalent to the signatures $\sigma$ defined in Sect. 1.6. We deduce that the eigenvectors of $S$ are the columns of $F$ and that they are associated with $\lambda_S = \pm 1$ eigenvalues. Using Equation (1.16), we can determine that the eigenspaces associated with both eigenvalues have the same dimension:

$$\dim E^S_{+1} = \dim E^S_{-1} = \frac{N_e}{2}. \tag{2.13}$$

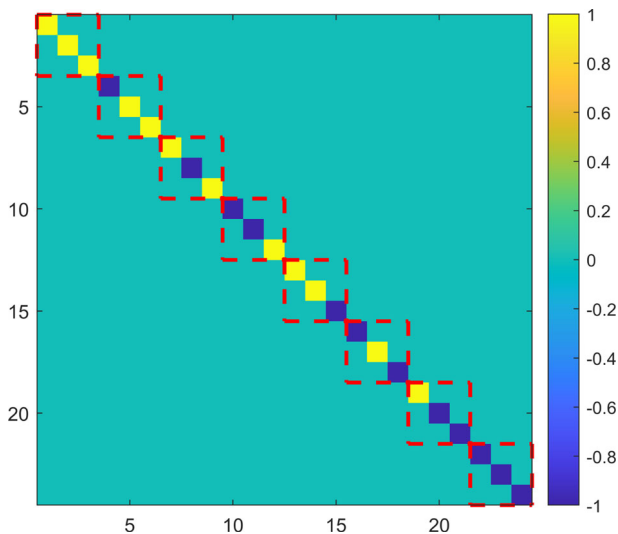

**Fig. 5** The diagonalized shift operator $\hat{S}$ in $\mathcal{H}^\mathcal{S} \otimes \mathcal{H}^\mathcal{C}$ for $n = 3$. The $\hat{S}(p)$ blocks are delimited by the dashed lines

We set $\hat{S}(p)$ to be the $n \times n$ block indexed by $p$. A quick way to find $\hat{S}(p)$ is to map the binary elements of $\rho$, the binary representation of $p$, to $+1/-1$ as noted before, then to place those elements on the diagonal.

### 2.2 The coin operator

The coin operator $C$ is a uniform diffusion operator in $\mathcal{H}^\mathcal{C}$. Its structure is based on the $n \times n$ Grover diffusion operator $G$ defined by

$$G = -I_n + 2|u\rangle\langle u|. \tag{2.14}$$

The coin operator only affects $\mathcal{H}^\mathcal{C}$. Therefore, its representation in $\mathcal{H}^\mathcal{S} \otimes \mathcal{H}^\mathcal{C}$ is $C = I_N \otimes G$, which gives a block diagonal matrix as seen in Fig. 6.

The eigendecompositions of $G$ and $C$ are easily linked. Let us denote those decompositions $G = V_G D_G V_G^\dagger$ and $C = V_C D_C V_C^\dagger$. Then,

$$C = I_N \otimes G, \tag{2.15}$$
$$= \left(I_N I_N I_N^\dagger\right) \otimes \left(V_G D_G V_G^\dagger\right), \tag{2.16}$$
$$= (I_N \otimes V_G)(I_N \otimes D_G)(I_N \otimes V_G)^\dagger, \tag{2.17}$$

and so $V_C = I_N \otimes V_G$ and $D_C = I_N \otimes D_G$.

Observe that $G|u\rangle = |u\rangle$, therefore $|u\rangle$ is an eigenvector of G associated with the eigenvalue $+1$. Also, $G\Lambda = -\Lambda$ (the $n \times (n-1)$ matrix $\Lambda$ is defined in Sect. 1.1 as the kernel of $\langle u|$); therefore, the $n-1$ columns of $\Lambda$ are all eigenvectors associated with the eigenvalue $-1$. The dimensions of the eigenspaces of $G$ are $\dim E^G_{+1} = 1$ and $\dim E^G_{-1} = n - 1$, and so the dimensions of the eigenspaces of $C$ are:

$$\dim E^C_{+1} = N, \tag{2.18}$$
$$\dim E^C_{-1} = N_e - N. \tag{2.19}$$

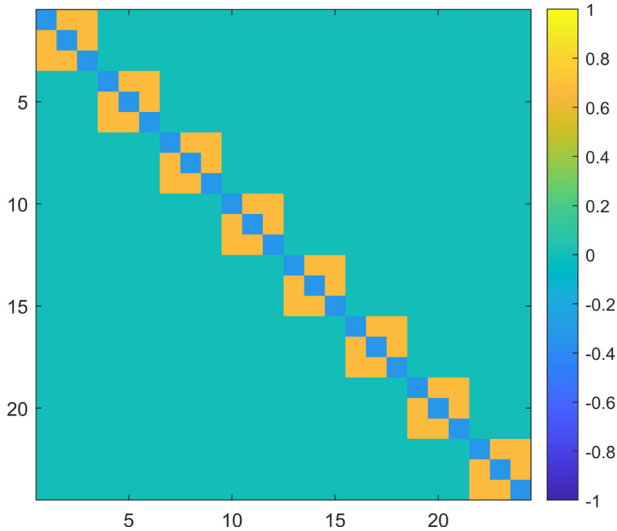

**Fig. 6** The coin operator in $\mathcal{H}^S \otimes \mathcal{H}^C$ for $n = 3$

A useful observation is that $C$ is invariant under the Fourier transform. We have

$$FCF = (H_N \otimes I_n)(I_N \otimes G)(H_N \otimes I_n), \tag{2.20}$$
$$= (H_N I_N H_N) \otimes (I_n G I_n), \tag{2.21}$$
$$= I_N \otimes G, \tag{2.22}$$
$$= C. \tag{2.23}$$

## 2.3 The oracle

The oracle $O$ is the core of the quantum search algorithm. Its structure varies according to the number of solutions and their positions on the hypercube. The easiest way to design the oracle operator is to create a diagonal matrix with $-1$ at the solution positions and $+1$ elsewhere, but we use another structure which gives a smaller eigenspace $E^O_-$. As the problem is to find the correct positions on the hypercube, all directions associated with a same vertex are equally valid. Therefore, the solutions always come in groups of $n$ in $\mathcal{H}^S \otimes \mathcal{H}^C$ if we work with a block diagonal oracle. If an $n \times n$ block corresponds to a group of solutions, it is set to $-G$, where $G$ is the Grover diffusion operator defined in Sect. 2.2. Else, the block is set to $I_n$. An example of oracle for a $n = 3$ qubit problem with solutions at position $p = 3$ and $p = 6$ is given in Fig. 7. In this example, the Grover diffusion operator is

$$G = -I_3 + 2\,|u_3\rangle\langle u_3| = -\begin{bmatrix} 1 & 0 & 0 \\ 0 & 1 & 0 \\ 0 & 0 & 1 \end{bmatrix} + \frac{2}{3}\begin{bmatrix} 1 & 1 & 1 \\ 1 & 1 & 1 \\ 1 & 1 & 1 \end{bmatrix} = \frac{1}{3}\begin{bmatrix} -1 & 2 & 2 \\ 2 & -1 & 2 \\ 2 & 2 & -1 \end{bmatrix}. \quad (2.24)$$

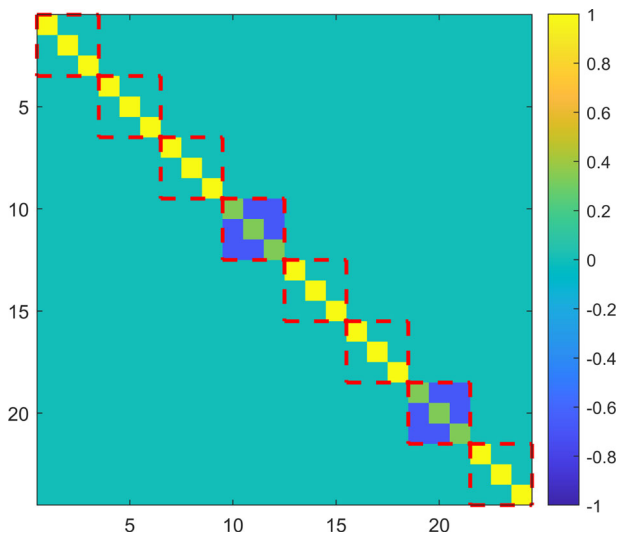

**Fig. 7** The oracle for $n = 3$ with solutions at positions 3 and 6. The blocks corresponding to the $N = 8$ outputs are delimited by the dashed lines

We can see that the $M = 2$ blocks corresponding to the solutions in position $p = 3$ and $p = 6$ are set to $-G$, and that the $N - M = 6$ others are set to $I$.

As we already know the eigendecomposition of $G$, we can immediately deduce the dimensions of the oracle eigenspaces and their eigenvalues:

$$\dim E_+^O = N_e - M, \quad (2.25)$$
$$\dim E_-^O = M. \quad (2.26)$$

The eigenvectors associated with $-1$ are $|1_p\rangle \otimes |u_n\rangle$, where $p$ is the position of a solution. It is interesting to note that $E_-^O$ is the subspace of the solution space $E^s$ corresponding to the solutions uniformly spread out in all directions and that $E_+^O$ is, therefore, the union of the non-solution space $E^{\bar{s}}$ and the non-uniformly spread out solutions. Note that $O$ and $C$ commute, as they are both block diagonal with only $I$ and $\pm G$ as possible blocks. Also, $G^2 = I_n$ so $O^2 = I_{N_E}$.

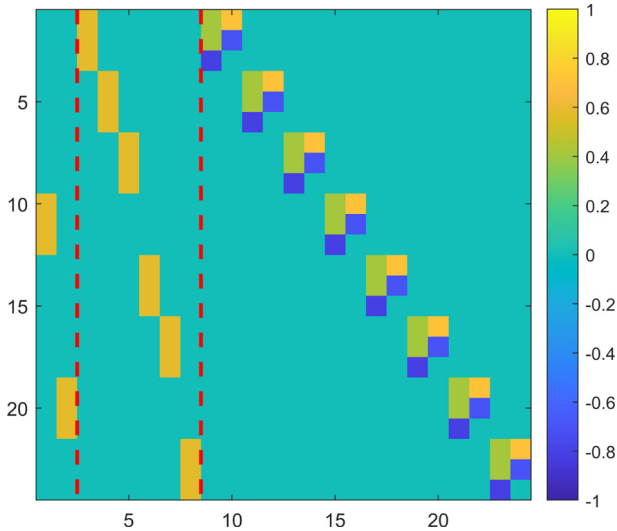

**Fig. 8** The generators concatenation $L_{1,2,3}$ for $n = 3$ with solutions at positions 2 and 5. The generators are delimited by the dashed lines

## 3 Generators

In this section, we will build "generators", matrices which will be useful in what follows. We define the first three generators as:

$$L_1 = I_N^s \otimes |u_n\rangle, \tag{3.1}$$

$$L_2 = I_N^{\bar{s}} \otimes |u_n\rangle, \tag{3.2}$$

$$L_3 = I_N \otimes \Lambda, \tag{3.3}$$

where $I_N^s$ and $I_N^{\bar{s}}$ are submatrices defined in Sect. 1.7.

These matrices have orthonormal columns and taken globally; they constitute an orthonormal basis of $\mathcal{H}$. We will see in the next section that they generate subspaces which are closely related to the eigenspaces of the quantum walk operators. Their sizes are, respectively, $N_e \times M$, $N_e \times (N - M)$ and $N_e \times (N_e - N)$. Their horizontal concatenation forms a unitary $N_e \times N_e$ matrix $L_{1,2,3}$ shown in Fig. 8. We will also use $L_{1,2}$ the horizontal concatenation of $L_1$ and $L_2$.

We can show that $L_{1,2}$ and $L_3$ span the same subspaces in the Fourier domain as in the original domain: we have $FL_3 = H_N \otimes \Lambda$. Since we can multiply on the right by any non-singular matrix without changing the spanned subspace, we can multiply by $H_N \otimes I_n$ and recover $L_3$ once more. Therefore,

$$F \operatorname{span}\{L_3\} = \operatorname{span}\{L_3\}. \tag{3.4}$$

The proof for $L_{1,2}$ is done the same way by replacing $\Lambda$ by $|u\rangle$.

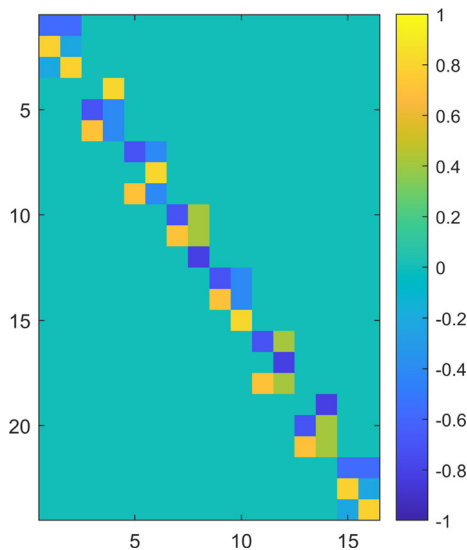

**Fig. 9** Generator $L'_3$ for $n = 3$

It is possible to create a variation of $L_3$ with most of its columns eigenvectors of $\hat{S}$. This variation will be denoted $L'_3$. To do so, we will replace each block individually by a custom block $\Lambda_p$. Those blocks all span the same subspace as $\Lambda$, as all their columns are orthogonal to $|u_n\rangle$. That means $L'_3$ spans the same subspace as $L_3$. Each $\Lambda_p$ is itself the concatenation of three matrices, $\Lambda_p^+$, $\Lambda_p^-$ and $\Lambda_p^\circ$, whose sizes depends on the Hamming weight $w$ of the binary representation of the $p$ indices:

- $\Lambda_p^-$ is an $n \times (w-1)$ matrix (or an empty matrix if $w \leq 1$). It has nonzero elements only at the $w$ rows with signature $\sigma = -1$. Therefore, its columns are eigenvectors of $\hat{S}$ associated with the eigenvalue $-1$. Those elements form $w - 1$ vectors of length $w$ all orthogonal to $|u_w\rangle$ and to each other.
- $\Lambda_p^+$ is an $n \times (n - w - 1)$ matrix (or an empty matrix if $n - w \leq 1$). It has nonzero elements only at the $n - w$ rows with signature $\sigma = +1$. Therefore, its columns are eigenvectors of $\hat{S}$ associated with the eigenvalue $+1$. Those elements form $n - w - 1$ vectors of length $n - w$ all orthogonal to $|u_{n-w}\rangle$ and to each other.
- $\Lambda_p^\circ$ is a column vector if $w \neq 0$ and $w \neq n$, and otherwise is empty. Its elements are

$$\sqrt{(n-w)/(nw)} \text{ where } \sigma = -1, \quad (3.5)$$

$$-\sqrt{k/(n(n-w))} \text{ where } \sigma = +1. \quad (3.6)$$

This column is not an eigenvector of $\hat{S}$, which is to be expected considering all of them are either in $\Lambda_p^-$ or $\Lambda_p^+$.

The resulting matrix $L'_3$ is shown in Fig. 9.

We define $L'^-_3$, $L'^+_3$ and $L'^\circ_3$ as the submatrices of $L'_3$ that are made up of the columns corresponding, respectively, to the $\Lambda_p^-$, $\Lambda_p^+$ and $\Lambda_p^\circ$. The size of $L'^-_3$ and $L'^+_3$ is $N_e \times (N_e/2 - N + 1)$, and the size of $L'^\circ_3$ is $N_e \times (N - 2)$.

**Table 1** Elementary operator eigenspaces analysis

| Eigenspace | Generator | Dimension |
|---|---|---|
| $E_\pm^S$ | $F^\pm$ | $\frac{N_e}{2}$ |
| $E_-^C$ | $L_3$ | $N_e - N$ |
| $E_+^C$ | $L_{1,2}$ | $N$ |
| $E_-^O$ | $L_1$ | $M$ |
| $E_+^O$ | $L_{2,3}$ | $N_e - M$ |
| $E_{-,+}^{C,O}$ | $L_3$ | $N_e - N$ |
| $E_{+,-}^{C,O}$ | $L_1$ | $M$ |
| $E_{+,+}^{C,O}$ | $L_2$ | $N - M$ |
| $E_-^{CO}$ | $L_{1,3}$ | $N_e - N + M$ |
| $E_+^{CO}$ | $L_2$ | $N - M$ |
| $E_{\pm,-}^{S,C}$ | $FL_3'^\pm$ | $\frac{N_e}{2} - N + 1$ |
| $E_{-,+}^{S,C}$ | $|l_-\rangle$ | 1 |
| $E_{+,+}^{S,C}$ | $|u\rangle$ | 1 |
| $E_{\pm,-,+}^{S,C,O}$ | $FL_3'^\pm$ | $\frac{N_e}{2} - N + 1$ |
| $E_{\pm,-}^{S,CO}$ | $L_{1,3} \perp F^\mp$ | $\frac{N_e}{2} - N + M$ |

$\perp F^\mp$ denotes a constraint of orthogonality

We also define the two generators $F^+$ and $F^-$ as the submatrices of $F$ that correspond, respectively, to the columns with signature $\sigma = +1$ and $\sigma = -1$. Both have size $N_e \times N_e/2$.

Finally, we define $|l_-\rangle$ as the Fourier transform of the column of $L_{1,2}$, which has all of its nonzero elements having a signature $\sigma = -1$, that is $|1_{N-1}\rangle \otimes |u_n\rangle$. Therefore,

$$|l_-\rangle = F\left(|1_{N-1}\rangle \otimes |u_n\rangle\right), \quad (3.7)$$

where $|1_{N-1}\rangle$ is a vector having a 1 at the position $N-1$ and 0 elsewhere, as defined in Sect. 1.1.

## 4 Eigenspaces junctions

In this section, we propose an exhaustive analysis of the joint eigenspaces of the quantum walk operators $S$, $C$ and $O$ in order to study the eigendecompositions of $U$ and $Q$. All these results can be found in Table 1. While some proofs are trivial, others are quite tedious and the reader may skip these and proceed without difficulty from the results in the table.

We already determined the dimension of the eigenspaces of the three elementary operators, and we can easily find that they are all spanned by one of the generators defined in Sect. 3. For instance, $E_-^O = \text{span}\{L_1\}$ and $E_+^C = \text{span}\{L_{1,2}\}$.

In the study of the joint eigenspaces, some cases are easily resolved. For instance, we have

$$E_{-,+}^{C,O} = E_-^C \cap E_+^O, \tag{4.1}$$
$$= \text{span}\{L_3\} \cap \text{span}\{L_{2,3}\}, \tag{4.2}$$
$$= \text{span}\{L_3\}, \tag{4.3}$$
$$= E_-^C. \tag{4.4}$$

Some of the joint eigenspaces do not appear in the table because they are empty:

$$E_{-,-}^{C,O} = E_-^C \cap E_-^O, \tag{4.5}$$
$$= \text{span}\{L_3\} \cap \text{span}\{L_1\}, \tag{4.6}$$
$$= \emptyset. \tag{4.7}$$

A useful step for what follows is to work with the operator $CO$. It is straightforward because $C$ and $O$ commute, which means they share the same eigenspaces. Therefore, we have

$$E_-^{CO} = E_{-,+}^{C,O} \cup E_{+,-}^{C,O}, \tag{4.8}$$
$$= \text{span}\{L_3\} \cup \text{span}\{L_1\}, \tag{4.9}$$
$$= \text{span}\{L_{1,3}\}, \tag{4.10}$$

and

$$E_+^{CO} = E_{-,-}^{C,O} \cup E_{+,+}^{C,O}, \tag{4.11}$$
$$= E_{+,+}^{C,O}, \tag{4.12}$$
$$= \text{span}\{L_2\}, \tag{4.13}$$

because $E_{-,-}^{C,O} = \emptyset$.

We begin the analysis of the joint eigenspaces of $S$ and $C$ with the observation that $C$ is invariant under the Fourier transform, as seen in Sect. 2.2, which means we can search for $E_{\pm,\pm}^{\hat{S},C}$ then switch back to the original domain. We have also seen that a vector is in $E_{\pm}^{\hat{S}}$ if it has nonzero elements only in positions with signature $\pm 1$. Since $E_-^C = \text{span}\{L_3\} = \text{span}\{L_3'\}$, we have

$$E_{\pm,-}^{\hat{S},C} = \text{span}\{F^\pm\} \cap \text{span}\{L_3'\}, \tag{4.14}$$
$$= \text{span}\{L_3'^\pm\}, \tag{4.15}$$

and so,

$$E_{\pm,-}^{S,C} = F\,\text{span}\{L_3'^\pm\} \tag{4.16}$$

Therefore, the dimension of $E_{\pm,-}^{S,C}$ is:

$$\dim E_{\pm,-}^{S,C} = \frac{N_e}{2} - N + 1. \qquad (4.17)$$

The next step is to note that $E_+^C = \text{span}\{L_{1,2}\}$ and that $\text{span}\{L_{1,2}\} = F\,\text{span}\{L_{1,2}\}$. There is only one column of $L_{1,2}$ in $E_-^{\hat{S}}$, the one that has all its elements with signature $-1$. We defined its Fourier transform as $|l_-\rangle$ above. Similarly the only column of $L_{1,2}$ in $E_+^{\hat{S}}$ is the one that has nonzero elements only in positions with signature $+1$. The Fourier transform of such a column is always the uniform vector $|u\rangle$.

$$E_{-,+}^{S,C} = \text{span}\{|l_-\rangle\}, \qquad (4.18)$$
$$E_{+,+}^{S,C} = \text{span}\{|u\rangle\}. \qquad (4.19)$$

Obviously, they both have dimension 1.

The only eigenspace that $C$ and $O$ share together that intersects $E_\pm^S$ is $\text{span}\{L_3\}$. We have

$$E_{\pm,-,+}^{S,C,O} = E_\pm^S \cap E_{-,+}^{C,O}, \qquad (4.20)$$
$$= E_\pm^S \cap E_-^C, \qquad (4.21)$$
$$= E_{\pm,-}^{S,C}. \qquad (4.22)$$

Therefore, $E_{\pm,-,+}^{S,C,O}$ is equal to $E_{\pm,-}^{S,C}$, while all other joint eigenspaces of $S$, $C$ and $O$ are empty.

We note that $E_{+,-}^{S,CO}$ includes $E_{+,-,+}^{S,C,O} = E_{+,-}^{S,C}$, which is spanned by $FL_3'^+$. We will create eigenvectors that span the complement of $E_{+,-}^{S,C}$ in $E_{+,-}^{S,CO}$. In the Fourier domain, $L_3'^+$ spans $E_{+,-}^{S,C}$ and $L_3'^-$ is orthogonal to $E_+^S$. Then, we will build those eigenvectors $|\varepsilon\rangle$ as linear combinations of the columns of $FL_1$ and $F_3'^o$. We have

$$|\varepsilon\rangle = FL_1|\varepsilon_1\rangle + F_3'^o|\varepsilon_3\rangle, \qquad (4.23)$$

where $|\varepsilon_1\rangle$ and $|\varepsilon_3\rangle$ are vectors of length $M$ and $N-2$. We also observe that the first and last rows of $F_3'^o$ are zero because there is no $\Lambda_p^o$ for $p = 0$ and $p = N-1$ and because

$$FL_1|\varepsilon_1\rangle = \left(H_N^s \otimes |u_n\rangle\right)|\varepsilon_1\rangle, \qquad (4.24)$$
$$= \left(H_N^s|\varepsilon_1\rangle\right) \otimes |u_n\rangle. \qquad (4.25)$$

In order to obtain $|\varepsilon\rangle$ in $E_+^{\hat{S}}$, we must cancel the elements with signature $-1$. All possible $|\varepsilon\rangle$ have to cancel the last block because all its elements always have signature $-1$. Since, as already noted, $L_3'^o$ is already zero in this block, we only have to cancel the $|\varepsilon_1\rangle$ component. Let us define $\langle h|$ as the last row of $H_N^s$. We have

$$\langle h| = \frac{1}{\sqrt{N}} \left[ (-1)^{w_1} \cdots (-1)^{w_M} \right], \quad (4.26)$$

where $w_k$ is the weight of the binary index of the $k$-th solution. We must have $\langle h|\varepsilon_1\rangle = 0$. Since $\langle h|$ is a $M$ length vector, we can find $M-1$ orthogonal $|\varepsilon_1\rangle$. We also can show that if $|\varepsilon_1\rangle$ is known, then we can determine a unique $|\varepsilon_3\rangle$. Let us consider any column $\Lambda_p^o$ of $L_3'^o$. Inside each block $p$, the positions with signature $-1$ contain $\sqrt{(n-w)/(nw)}$ and $FL_1|\varepsilon_1\rangle$ contains a value indexed by $p$ of $(H_N^s|\varepsilon_1\rangle)/\sqrt{n}$, denoted $(H_N^s|\varepsilon_1\rangle)_p/\sqrt{n}$. Then, the part of $|\varepsilon_3\rangle$ associated with block $p$ guarantees:

$$\frac{1}{\sqrt{n}} \left( H_N^s|\varepsilon_1\rangle \right)_p + \sqrt{\frac{n-w}{nw}} |\varepsilon_3(p)\rangle = 0. \quad (4.27)$$

Therefore,

$$|\varepsilon_3(p)\rangle = -\sqrt{\frac{w}{n-w}} \left( H_N^s|\varepsilon_1\rangle \right)_p. \quad (4.28)$$

We can deduce that the dimension of the complement of $E_{-,-}^{S,C}$ in $E_{-,-}^{S,CO}$ is $M-1$, and since $\dim E_{-,-}^{S,C} = N_e/2 - N + 1$, we have

$$\dim E_{-,-}^{S,CO} = \frac{N_e}{2} - N + M. \quad (4.29)$$

The proof is similar for $E_{+,-}^{S,CO}$. In that case, we define $\langle h|$ as the first row of $H_N^s$, that is $\langle u|$.

$$\dim E_{+,-}^{S,CO} = \frac{N_e}{2} - N + M. \quad (4.30)$$

We can also see that $E_{+,-}^{S,CO}$ includes $E_{+,-,+}^{S,C,O} = E_{+,-}^{S,C}$, which is spanned by $FL_3'^+$

## 5 Eigendecomposition of the uniform walk operator

In this section, we will determine the total eigendecomposition of the uniform walk operator $U$. As in the last section, the reader can refer to Table 2 for a summary of the results.

The uniform walk operator is $U = SC$; therefore, we have

$$E_-^U \supseteq E_{+,-}^{S,C} \cup E_{-,+}^{S,C}, \quad (5.1)$$
$$E_+^U \supseteq E_{-,-}^{S,C} \cup E_{+,+}^{S,C}, \quad (5.2)$$

Table 2 Uniform walk operator eigendecomposition

| Eigenspace | Generator | Dimension |
|---|---|---|
| $E_-^U$ | $[FL_3'^+ \ \|l_-\rangle]$ | $\frac{N_e}{2} - N + 2$ |
| $E_+^U$ | $[FL_3'^- \ \|u\rangle]$ | $\frac{N_e}{2} - N + 2$ |
| $E_{\lambda_w}^U$ | $V_w$ | $\binom{n}{w}$ |
| $E_{\lambda_w^*}^U$ | $V_w^*$ | $\binom{n}{w}$ |

$[A \ B]$ denotes the horizontal concatenation of $A$ and $B$

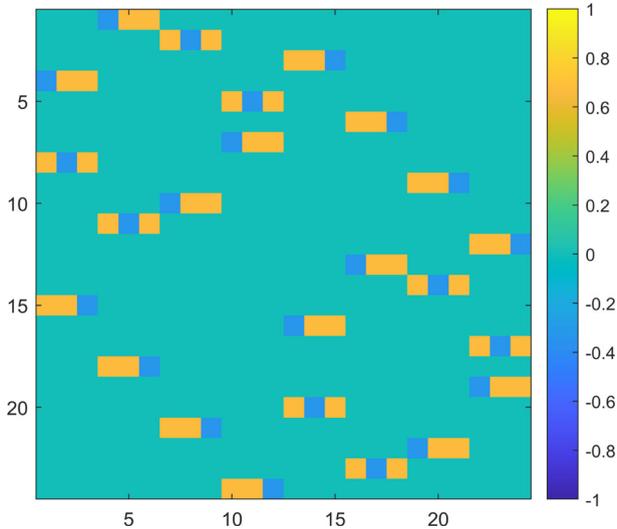

**Fig. 10** The uniform walk operator for $n = 3$

which implies:

$$\dim E_-^U \geq \frac{N_e}{2} - N + 2, \quad (5.3)$$

$$\dim E_+^U \geq \frac{N_e}{2} - N + 2. \quad (5.4)$$

As illustrated in Fig. 10, the structure of $U$ is complicated and turns out to be much simple in the Fourier domain:

$$\hat{U} = FUF, \quad (5.5)$$
$$= (FSF)(FCF), \quad (5.6)$$
$$= \hat{S}C. \quad (5.7)$$

We can see that $\hat{U}$ is a block diagonal matrix containing $N$ blocks $\hat{U}_p$ such as

$$\hat{U}_p = \hat{S}(p)G, \quad (5.8)$$

where $\hat{S}(p)$ is the block of $\hat{S}$ associated with $p$ defined in Sect. 2.1, and $G$ the Grover diffusion operator defined in Sect. 2.2. The diagonal of each $\hat{S}(p)$ accounts for $n - w$ times '+1' and $w$ times '−1'.

The only coefficient of the diagonal of $G$ is $-1 + 2/n$; therefore, the trace of each $\hat{U}_p$ is

$$\mathrm{tr}\left(\hat{U}_p\right) = (((n-w) - w)\left(-1 + \frac{2}{n}\right), \tag{5.9}$$

$$= (n - 2w)\left(-1 + \frac{2}{n}\right), \tag{5.10}$$

$$= -n + 2w + 2\left(1 - 2\frac{w}{n}\right). \tag{5.11}$$

For any matrix, the sum of its eigenvalues is equal to the trace, and we already know that when $p \neq 0$ and $p \neq N-1$, there are $(n - w - 1)$ '−1' eigenvalues and $(w - 1)$ '+1' eigenvalues. The sum of these $n - 2$ eigenvalues is therefore $-n + 2w$, and there are still two eigenvalues left to determine, which we will denote by $\lambda_w$ and $\lambda_w^*$. Their sum must be $2(1 - 2(w/n))$, which means:

$$\mathrm{Re}\{\lambda_w\} = 1 - 2\frac{w}{n}. \tag{5.12}$$

Since $U$ is unitary, so are the $\hat{U}_p$ and $|\lambda_w| = 1$. Thus,

$$\lambda_w = 1 - 2\frac{w}{n} + \frac{2\mathrm{i}}{n}\sqrt{w(n-w)}. \tag{5.13}$$

Note that $\lambda_w$ and $\lambda_w^*$ are complex conjugate eigenvalues. The choice of which one is $\lambda_w$ is arbitrary.

We define the vector $|v_w\rangle$ by:

$$|v_w\rangle = \frac{1}{\sqrt{2w}}|\rho\rangle - \frac{\mathrm{i}}{\sqrt{2(n-w)}}|\bar{\rho}\rangle, \tag{5.14}$$

where $|\rho\rangle$ and $|\bar{\rho}\rangle$ are the vectors containing the binary representation of the block index and its negation. We will prove that $|v_w\rangle$ is an eigenvector of $\hat{U}_p$ associated with the eigenvalue $\lambda_w$. Note that

$$\langle u_n | v_w \rangle = \frac{1}{2}\sqrt{\frac{nw}{2}}(1 - \lambda_w), \tag{5.15}$$

since it is the sum of all terms in $|v_w\rangle$ divided by $\sqrt{n}$. We also have

$$\hat{S}(p)|v_w\rangle = -|v_w\rangle^* \tag{5.16}$$

and

$$\hat{S}(p)|u_n\rangle = \frac{|\bar{\rho}\rangle - |\rho\rangle}{\sqrt{n}}. \quad (5.17)$$

We deduce that

$$\hat{U}_p|v_w\rangle = \hat{S}(p)G|v_w\rangle, \quad (5.18)$$
$$= \hat{S}(p)\left(-I_n + 2|u_n\rangle\langle u_n|\right)|v_w\rangle, \quad (5.19)$$
$$= \left(-\hat{S}(p) + \frac{2}{\sqrt{n}}(|\bar{\rho}\rangle - |\rho\rangle)\langle u_n|\right)|v_w\rangle, \quad (5.20)$$
$$= -\hat{S}(p)|v_w\rangle + \frac{2}{\sqrt{n}}\langle u_n|v_w\rangle(|\bar{\rho}\rangle - |\rho\rangle), \quad (5.21)$$
$$= \frac{1}{\sqrt{2w}}|\rho\rangle + \frac{i}{\sqrt{2(n-w)}}|\bar{\rho}\rangle + \frac{1}{n}\left(2\sqrt{2w} - i\sqrt{2(nw)}\right)(|\bar{\rho}\rangle - |\rho\rangle), \quad (5.22)$$
$$= \left(1 - \frac{2w}{n} + \frac{2i}{n}\sqrt{w(n-w)}\right)\left(\frac{1}{\sqrt{2w}}|\rho\rangle - \frac{i}{\sqrt{2(nw)}}|\bar{\rho}\rangle\right), \quad (5.23)$$
$$= \lambda_w|v_w\rangle. \quad (5.24)$$

Thus, $|v_w\rangle$ and $|v_w\rangle^*$ constitute $2(N-2)$ eigenvectors. Taking into consideration the $2(N_e/2 - N + 2)$ already known from $E_{\pm,\pm}^{S,C}$, this accounts for all of them. Since none of the $2(N-2)$ new eigenvectors are associated with an eigenvalue $\pm 1$, we have

$$E_-^U = E_{+,-}^{S,C} \cup E_{-,+}^{S,C}, \quad (5.25)$$
$$E_+^U = E_{-,-}^{S,C} \cup E_{+,+}^{S,C}, \quad (5.26)$$

which implies

$$\dim E_-^U = \frac{N_e}{2} - N + 2, \quad (5.27)$$
$$\dim E_+^U = \frac{N_e}{2} - N + 2. \quad (5.28)$$

We can also determine the size of each $E_{\lambda_w}^U$, as there are $\binom{n}{w}$ positions whose binary index weight is $w$. We have

$$\dim E_{\lambda_w}^U = \binom{n}{w}. \quad (5.29)$$

To summarize, we can list the eigenvectors of $U$ in the Fourier domain, i.e., the eigenvectors of $\hat{U}$:

- The orthonormal basis of $E_-^{\hat{U}}$ is composed of:
  - The $N_e/2 - N + 1$ columns $L_3'^+$.
  - The column of $L_{1,2}\,|1_{N-1}\rangle \otimes |u_n\rangle$.

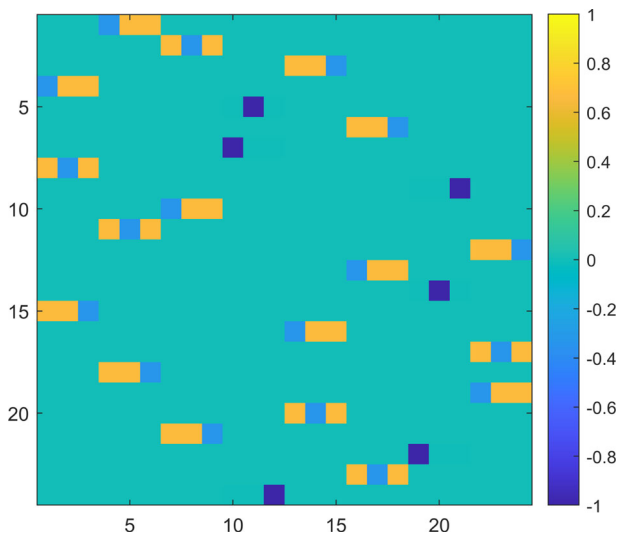

**Fig. 11** The effective walk operator for $n = 3$ with solutions at positions 3 and 6

- The orthonormal basis of $E_+^{\hat{U}}$ is composed of:
  - The $N_e/2 - N + 1$ columns $L_3'^{-}$.
  - The column of $L_{1,2} |1_0\rangle \otimes |u_n\rangle$.
- The orthonormal basis of $E_{\lambda_w}^{\hat{U}}$ is composed of the $\binom{n}{w}$ vectors $|1_p\rangle \otimes |v_w\rangle$.
- The orthonormal basis of $E_{\lambda_w^*}^{\hat{U}}$ is composed of the $\binom{n}{w}$ vectors $|1_p\rangle \otimes |v_w\rangle^*$ for all $p$ whose binary representation Hamming weight is $w$.

If needed, the eigenvectors of $U$ can easily be deduced from those of $\hat{U}$ using a multiplication by $F$ to switch back to the original domain. We define two new conjugate generators $V_w$ and $V_w^*$ that span, respectively, $E_{\lambda_w}^{\hat{U}}$ and $E_{\lambda_w^*}^{\hat{U}}$. Their respective columns are $F|1_p\rangle \otimes |v_w\rangle$ and $F|1_p\rangle \otimes |v_w\rangle^*$.

## 6 Dimension of the complement space

As explained in the introduction, the purpose of the eigenspaces study is to determine the dimension of $\mathcal{E}$ and the eigenvalues of the effective quantum walk operator $Q$ associated with it. Since $Q = UO$, it has a complicated structure in both original and Fourier domains, and we will need to use results from previous sections. Its structure in the original domain is shown in Fig. 11. Once again, we will summarize the results in Table 3 at the end of this section.

Let us start by demonstrating that $E_{\lambda,-}^{U,O} = \emptyset$ for any $\lambda$. We know that $E_-^O \subset E_+^C$, so

$$E_{\lambda,-}^{U,O} \subset E_{\lambda,+}^{U,C}. \tag{6.1}$$

**Table 3** Effective walk operator eigendecomposition

| Eigenspace | Generator | Dimension |
| --- | --- | --- |
| $E_{\pm,+}^{U,O}$ | $FL_3^{\prime\mp}$ | $\frac{N_e}{2} - N + 1$ |
| $E_{\lambda_w,+}^{U,O}$ | $F \ker\left\{(H_N^{s,w})^T\right\}$ | $\binom{n}{w} - r_w$ |
| $E_{\lambda_w^*,+}^{U,O}$ | $F \ker\left\{(H_N^{s,w})^T\right\}$ | $\binom{n}{w} - r_w$ |

$r_w$ denotes the rank of $H_N^{s,w}$

Furthermore, since $U = SC$, $E_{\lambda,+}^{U,C} = E_{\lambda,+}^{S,C}$, and

$$E_{\lambda,-}^{U,O} \subset E_{\lambda,+}^{S,C}, \qquad (6.2)$$

which means that $\lambda = \pm 1$, since $S$ does not have other eigenvalue. Therefore,

$$E_{-,-}^{U,O} \subset E_{-,+}^{S,C}, \qquad (6.3)$$
$$E_{+,-}^{U,O} \subset E_{+,+}^{S,C}, \qquad (6.4)$$

and so

$$E_{-,-}^{U,O} \subset \text{span}\{|l_-\rangle\}, \qquad (6.5)$$
$$E_{+,-}^{U,O} \subset \text{span}\{|u\rangle\}. \qquad (6.6)$$

Since both $|l_-\rangle$ and $|u\rangle$ contain only nonzero elements, they are not in span $\{L_1\} = E_-^O$ and finally

$$E_{\lambda,-}^{U,O} = \emptyset. \qquad (6.7)$$

Let us now consider $E_{\pm,+}^{U,O}$. We saw that

$$E_-^U = E_{+,-}^{S,C} \cup E_{-,+}^{S,C}, \qquad (6.8)$$
$$E_+^U = E_{-,-}^{S,C} \cup E_{+,+}^{S,C}, \qquad (6.9)$$

and

$$E_{\pm,-}^{S,C} = \text{span}\{FL_3^{\prime\pm}\}, \qquad (6.10)$$
$$E_{-,+}^{S,C} = \text{span}\{|l_-\rangle\}, \qquad (6.11)$$
$$E_{+,+}^{S,C} = \text{span}\{|u\rangle\}. \qquad (6.12)$$

We can show that $|u\rangle$ and $|l_-\rangle$ are not in $E_+^O$ for the same reason as they are not in $E_-^O$. Furthermore, span $\{FL_3^{\prime+}\} \subset$ span $\{L_3\}$, and since $E_+^O =$ span $\{L_{2,3}\}$,

$$E_{\pm,+}^{U,O} = E_{\mp,-}^{S,C}, \qquad (6.13)$$

and
$$\dim E^{U,O}_{\pm,+} = \frac{N_e}{2} - N + 1. \tag{6.14}$$

A secondary but important result is that both $E^{S,C}_{-,+}$ and $E^{S,C}_{-,+}$ are in $\mathcal{E}$, as they have no intersection with any eigenspace of $O$. That means $|u_{N_e}\rangle$ is in $\mathcal{E}$, and any quantum walk initialized in this state will remain in $\mathcal{E}$.

Note that $E^O_+ = \text{span}\{L_{2,3}\} = \ker\{L_1^T\}$. Denote by $\hat{O}$ the Fourier transform of $O$. Then,
$$E^{\hat{O}}_+ = \ker\left\{(FL_1)^T\right\}, \tag{6.15}$$

and
$$FL_1 = (H_N \otimes I_n)\left(I^s_N \otimes |u_n\rangle\right), \tag{6.16}$$
$$= H^s_N \otimes |u_n\rangle, \tag{6.17}$$

where $H^s_N$ is the submatrix defined in Sect. 1.7. It now follows that
$$E^{\hat{O}}_+ = \ker\left\{H^{s\,T}_N \otimes \langle u_n|\right\}. \tag{6.18}$$

In the Fourier domain, $E^{\hat{U}}_{\lambda_w} = \text{span}\{V_w\}$, where $V_w$ is the generator defined in Sect. 5. We have
$$E^{\hat{U},\hat{O}}_{\lambda_w,+} = \text{span}\{V_w\} \cap \ker\left\{H^{s\,T}_N \otimes \langle u_n|\right\}. \tag{6.19}$$

Since $V_w$ is an $N_e \times \binom{n}{w}$ matrix, each unit vector of span$\{V_w\}$ can be expressed as $V_w|\nu\rangle$, where $|\nu\rangle$ is a unit vector of length $\binom{n}{w}$. If in addition
$$\left(H^{s\,T}_N \otimes \langle u_n|\right)(V_w|\nu\rangle) = 0, \tag{6.20}$$

such a vector also belongs to $E^{\hat{O}}_+$, which means $|\nu\rangle$ must be orthogonal with all of the $\left(H^{s\,T}_N \otimes \langle u_n|\right) V_w$ columns. These columns are:
$$\left(H^{s\,T}_N \otimes \langle u_n|\right)(|1_p\rangle \otimes |v_w\rangle), \tag{6.21}$$
$$= \langle u_n|v_w\rangle \left(H^s_N\right)^T |1_p\rangle. \tag{6.22}$$

We identify here $H^{s\,T}_N|1_p\rangle$ as the column of $H^{s\,T}_N$ indexed by $p$ and $\langle u_n|v_w\rangle$ as a scalar value that we will denote by $\alpha_w$. We have
$$\alpha_w = \sqrt{\frac{w}{2}} - i\sqrt{\frac{n-w}{2}}, \tag{6.23}$$

which means
$$\left(H^{s\,T}_N \otimes \langle u_n|\right) V_w = \alpha_w H^{s,w\,T}_N. \tag{6.24}$$

Since $\alpha_w \neq 0$ for any $w$, we have

$$H_N^{s,w\mathrm{T}} |\nu\rangle = 0, \tag{6.25}$$

that is

$$|\nu\rangle \in \ker\left\{H_N^{s,w\mathrm{T}}\right\}, \tag{6.26}$$

and finally

$$\dim E_{\lambda_w,+}^{U,O} = \dim\left(\ker\left\{H_N^{s,w\mathrm{T}}\right\}\right), \tag{6.27}$$

$$= \binom{n}{w} - \mathrm{rank}\left\{H_N^{s,w}\right\}. \tag{6.28}$$

In the following, we will denote $\mathrm{rank}\left\{H_N^{s,w}\right\}$ by $r_w$. The same proof applies to $V_w^*$, so

$$\dim E_{\lambda_w^*,+}^{U,O} = \binom{n}{w} - r_w. \tag{6.29}$$

We now have everything we need to determine the dimension of $\overline{\mathcal{E}}$ and $\mathcal{E}$. We recall that

$$\overline{\mathcal{E}} = \bigcup_{\lambda_u, \lambda_o} E_{\lambda_u, \lambda_o}^{U,O}, \tag{6.30}$$

where $\lambda_u$ and $\lambda_o$ are the eigenvalues of $U$ and $O$. Therefore,

$$\dim \overline{\mathcal{E}} = 2\left(\frac{N_e}{2} - N + 1\right) + 2\sum_{w=1}^{n-1}\left(\binom{n}{w} - r_w\right), \tag{6.31}$$

$$= N_e - 2N + 2 + 2(N-2) - 2\sum_{w=1}^{n-1} r_w, \tag{6.32}$$

$$= N_e - 2 - 2\sum_{w=1}^{n-1} r_w. \tag{6.33}$$

Since $\sum_{w=0}^{n}\binom{n}{w} = 2^n$, we have:

$$\dim \overline{\mathcal{E}} = N_e - 2 - 2\sum_{w=1}^{n-1} r_w, \tag{6.34}$$

and finally

$$\dim \mathcal{E} = N_e - \dim \overline{\mathcal{E}}, \tag{6.35}$$

$$= 2 + 2\sum_{w=1}^{n-1} r_w. \tag{6.36}$$

We also know that $\overline{\mathcal{E}} \subseteq E_+^O$, so

$$\dim \overline{\mathcal{E}} \leq N_e - M. \tag{6.37}$$

Then, because $1 \leq r_w \leq \min(M, \binom{n}{w})$, we have:

$$\max(2n, M) \leq \dim \mathcal{E} \leq 2(n-1)M + 2. \tag{6.38}$$

We have shown that even if the dimension of the total space $\mathcal{H}$ increases exponentially with $n$, the dimension of $\mathcal{E}$ only grows linearly. For instance, for $n = 50$ and $M = 4$, $\dim \mathcal{H} \approx 5.6 \times 10^{16}$ while $\dim \mathcal{E} \leq 394$. This result does not imply that the quantum search algorithm can be run on a classical computer, but it allows us to compute efficiently the evolution of the probability of success.

## 7 Computation of the probability of success

### 7.1 Eigenvalues of the eigenspace of interest

Recall that $|u\rangle$ is the uniform state ($N_e$ dimensional in this case), $|s\rangle$ is the uniform superposition of all solutions and $|\bar{s}\rangle$ is the uniform superposition of all non-solutions. In the last section, we have established an upper bound on the dimension of $\mathcal{E}$ and shown that it includes $|u\rangle$ and $|l_-\rangle$. Since $E_-^O \subset \mathcal{E}$ and $E_-^O = \text{span}\{L_1\}$, we know that $|s\rangle \subset \mathcal{E}$. Then,

$$|\bar{s}\rangle = \sqrt{\frac{N_e}{N_e - M}}|u\rangle + \sqrt{\frac{M}{N_e - M}}|s\rangle, \tag{7.1}$$

so $|\bar{s}\rangle$ is a linear combination of $|u\rangle$ and $|s\rangle$ and is in $\mathcal{E}$ too. We also noted that if the initial state is in $\mathcal{E}$, then it will remain in $\mathcal{E}$ during the algorithm. We will therefore initiate with $|u\rangle$.

The goal of this section is to determine the eigenvalues of $Q$ associated with $\mathcal{E}$, as well as the components of $|u\rangle$ and $|s\rangle$ in this subspace. Then, we will see that it is easy to compute the probability of success for any number of iterations $t$.

If we use an orthonormal basis of eigenvectors of $Q$, we can represent the component of $Q$ in $\mathcal{E}$ as a $\dim \mathcal{E} \times \dim \mathcal{E}$ diagonal operator $Q_\mathcal{E}$, whose diagonal elements are its eigenvalues. If we denote those eigenvalues by $e^{i\varphi_k}$, the components of the state after $t$ iterations are:

$$\psi_t(k, l) = e^{i\varphi_k t} u(k, l), \tag{7.2}$$

and the probability of success

$$p_t = \left| \sum_{k,l} s(k, l)^* \psi_t(k, l) \right|^2, \tag{7.3}$$

where the terms $\psi_t(k, l)$, $u(k, l)$ and $s(k, l)$ are the respective components of $|\psi_t\rangle$, $|u\rangle$ and $|s\rangle$ in the eigenspace associated with $e^{i\varphi_k}$, and the parameter $l$ allows us to distinguish the components of a given eigenspace whose dimension is greater than 1. We also define $|u(k)\rangle$ and $|s(k)\rangle$ the vectors made of all $u(k, l)$ and $s(k, l)$ for a given $k$. The length of those vectors is the multiplicity of the $e^{i\varphi_k}$ eigenvalue.

Since $E_-^{CO} \cup E_+^{CO} = \mathcal{H}$, any eigenvector $|\varepsilon\rangle$ in $\mathcal{E}$ can be represented as the sum of two vectors belonging to $E_-^{CO}$ and $E_+^{CO}$:

$$|\varepsilon\rangle = |\varepsilon_-\rangle + |\varepsilon_+\rangle. \tag{7.4}$$

Let us denote $\lambda = e^{i\varphi}$ an eigenvalue of $Q$. We have

$$Q|\varepsilon\rangle = \lambda|\varepsilon\rangle = \lambda|\varepsilon_-\rangle + \lambda|\varepsilon_+\rangle, \tag{7.5}$$

and

$$Q|\varepsilon\rangle = SCO\left(|\varepsilon_-\rangle + |\varepsilon_+\rangle\right), \tag{7.6}$$
$$= -S|\varepsilon_-\rangle + S|\varepsilon_+\rangle. \tag{7.7}$$

Therefore,

$$-S|\varepsilon_-\rangle + S|\varepsilon_+\rangle = \lambda|\varepsilon_-\rangle + \lambda|\varepsilon_+\rangle. \tag{7.8}$$

Since $S^2 = I$, we have

$$-|\varepsilon_-\rangle + |\varepsilon_+\rangle = \lambda S|\varepsilon_-\rangle + \lambda S|\varepsilon_+\rangle. \tag{7.9}$$

Let us denote $P_\pm^S$ the projectors onto $E_\pm^S$. As seen in Sect. 1.5, if $S^2 = I$, then $P_\pm^S = (I \pm S)/2$. From the two last equations, we have

$$-P_+^S|\varepsilon_-\rangle + P_+^S|\varepsilon_+\rangle = \lambda P_+^S|\varepsilon_-\rangle + \lambda P_+^S|\varepsilon_+\rangle, \tag{7.10}$$
$$P_-^S|\varepsilon_-\rangle - P_-^S|\varepsilon_+\rangle = \lambda P_-^S|\varepsilon_-\rangle + \lambda P_-^S|\varepsilon_+\rangle, \tag{7.11}$$

that is

$$(1 - \lambda)P_+^S|\varepsilon_+\rangle = (1 + \lambda)P_+^S|\varepsilon_-\rangle, \tag{7.12}$$
$$(1 + \lambda)P_-^S|\varepsilon_+\rangle = (1 - \lambda)P_-^S|\varepsilon_-\rangle. \tag{7.13}$$

Suppose $\lambda \neq \pm 1$, the particular cases will be treated later. Since $\lambda = e^{i\varphi}$, we have

$$\frac{1 - \lambda}{1 + \lambda} = -\frac{e^{i\frac{\varphi}{2}} - e^{-i\frac{\varphi}{2}}}{e^{i\frac{\varphi}{2}} + e^{-i\frac{\varphi}{2}}} = -i \tan\frac{\varphi}{2}, \tag{7.14}$$

$$\frac{1 + \lambda}{1 - \lambda} = -\frac{1}{i \tan\frac{\varphi}{2}} = i \cot\frac{\varphi}{2}. \tag{7.15}$$

Then,

$$P_+^S|\varepsilon_+\rangle = i \cot\frac{\varphi}{2} P_+^S|\varepsilon_-\rangle, \tag{7.16}$$

$$P_-^S|\varepsilon_+\rangle = -\mathrm{i}\tan\frac{\varphi}{2}P_-^S|\varepsilon_-\rangle. \tag{7.17}$$

Because $P_+^S + P_-^S = I$, we can sum these two equations to obtain

$$|\varepsilon_+\rangle = \mathrm{i}\left(\cot\frac{\varphi}{2}P_+^S|\varepsilon_-\rangle - \tan\frac{\varphi}{2}P_-^S|\varepsilon_-\rangle\right), \tag{7.18}$$

and conversely

$$|\varepsilon_-\rangle = -\mathrm{i}\left(\tan\frac{\varphi}{2}P_+^S|\varepsilon_+\rangle - \cot\frac{\varphi}{2}P_-^S|\varepsilon_+\rangle\right), \tag{7.19}$$

Let us denote by $\hat{P}_+^S$ the Fourier transform of $P_+^S$. From the study of the eigenspaces of $S$ in Sect. 2.1, we know that $\hat{P}_+^S$ is a diagonal matrix which contains 1 at the positions with signature $+1$ and 0 elsewhere. In the same way, we define $\hat{P}_-^S$ the diagonal matrix which contains 1 at the positions with signature $-1$ and 0 elsewhere. Define

$$\hat{\Gamma}_\varphi = \cot\frac{\varphi}{2}\hat{P}_+^S - \tan\frac{\varphi}{2}\hat{P}_-^S, \tag{7.20}$$

so that

$$|\hat{\varepsilon}_+\rangle = \mathrm{i}\hat{\Gamma}_\varphi|\hat{\varepsilon}_-\rangle, \tag{7.21}$$
$$|\hat{\varepsilon}_-\rangle = -\mathrm{i}\hat{\Gamma}_\varphi^{-1}|\hat{\varepsilon}_+\rangle, \tag{7.22}$$

where $|\hat{\varepsilon}_+\rangle$ and $|\hat{\varepsilon}_-\rangle$ are the Fourier transforms of $|\varepsilon_+\rangle$ and $|\varepsilon_-\rangle$.

We have

$$|\varepsilon_+\rangle \in E_+^{CO} = \mathrm{span}\{L_2\}, \tag{7.23}$$
$$|\varepsilon_-\rangle \in E_-^{CO} = \mathrm{span}\{L_{1,3}\}, \tag{7.24}$$

but since $FL_3'^{\pm}$ span eigenspaces orthogonal to $\mathcal{E}$, we can restrict ourselves to $FL_3'^o$ instead of $L_3$. Then,

$$|\varepsilon_-\rangle = L_1|\varepsilon_1\rangle + FL_3'^o|\varepsilon_3\rangle, \tag{7.25}$$
$$|\varepsilon_+\rangle = L_2|\varepsilon_2\rangle. \tag{7.26}$$

Using the definitions of the generators in Eqs. (3.1) and (3.2), we have

$$|\varepsilon_-\rangle = I_N^s|\varepsilon_1\rangle \otimes |u_n\rangle + FL_3'^o|\varepsilon_3\rangle, \tag{7.27}$$
$$|\varepsilon_+\rangle = I_N^{\bar{s}}|\varepsilon_2\rangle \otimes |u_n\rangle. \tag{7.28}$$

Define the vectors

$$|e_-\rangle = I_N^s|\varepsilon_1\rangle, \tag{7.29}$$
$$|e_+\rangle = I_N^{\bar{s}}|\varepsilon_2\rangle. \tag{7.30}$$

Note that $FL_3'^{\circ}|\varepsilon_3\rangle$ is a vector whose per block average value is zero, therefore

$$|e_\pm\rangle = (I_N \otimes |u_n\rangle)|\varepsilon_\pm\rangle. \tag{7.31}$$

We can decompose $\hat{\Gamma}_\varphi^{-1}$ as

$$\hat{\Gamma}_\varphi^{-1} = \hat{D}_\varphi^{-1} \otimes I_n + \Upsilon_\varphi, \tag{7.32}$$

where $\hat{D}_\varphi^{-1}$ is a diagonal matrix whose diagonal elements are the per block average values of the diagonal of $\Gamma_\varphi^{-1}$

$$\hat{D}_\varphi^{-1}(p) = \left(1 - \frac{w_p}{n}\right)\tan\frac{\varphi}{2} - \frac{w_p}{n}\cot\frac{\varphi}{2}, \tag{7.33}$$

$w_p$ is the Hamming weight of the binary representation of $p$ and $\Upsilon_\varphi$ is another diagonal matrix whose per block average is zero. Let $|\hat{e}_\pm\rangle$ be the Fourier transform of $|e_\pm\rangle$. By using Equation (7.22), we have

$$(I_N \otimes \langle u_n|)|\hat{\varepsilon}_-\rangle = -\mathrm{i}\,(I_N \otimes \langle u_n|)\left(D_\varphi^{-1} \otimes I_n + \Upsilon_\varphi\right)(|\hat{e}_+\rangle \otimes |u_n\rangle), \tag{7.34}$$

$$|\hat{e}_-\rangle = -\mathrm{i}\,(I_N \otimes \langle u_n|)\left(D_\varphi^{-1} \otimes I_n\right)(|\hat{e}_+\rangle \otimes |u_n\rangle), \tag{7.35}$$

because $\Upsilon_\varphi(|\hat{e}_+\rangle \otimes |u_n\rangle) = 0$. Notice that

$$|\hat{e}_-\rangle = -\mathrm{i}\hat{D}_\varphi^{-1}|\hat{e}_+\rangle. \tag{7.36}$$

If the diagonal of $\hat{D}_\varphi^{-1}$ has no zero element (this case will be treated later), we have

$$|\hat{e}_+\rangle = \mathrm{i}\hat{D}_\varphi|\hat{e}_-\rangle, \tag{7.37}$$

were $\hat{D}_\varphi$ is a diagonal matrix whose elements are

$$\hat{D}_\varphi(p) = \left(\left(1 - \frac{w_p}{n}\right)\tan\frac{\varphi}{2} - \frac{w_p}{n}\cot\frac{\varphi}{2}\right)^{-1}. \tag{7.38}$$

Since they only depend on $w_p$, the elements of $\hat{D}_\varphi$ and its inverse can be indexed by the Hamming weight of their binary indices only. We will denote those elements $\hat{D}_\varphi(w_p)$.

Back to the original domain, we have

$$D_\varphi = H_N \hat{D}_\varphi H_N, \tag{7.39}$$

so

$$|e_+\rangle = i D_\varphi |e_-\rangle. \tag{7.40}$$

Define $D_\varphi^s$ to be the $M \times M$ submatrix of $D_\varphi$ that contains only the rows and columns associated with a solution. Then,

$$D_\varphi^s = H_N^{s\,\mathrm{T}} \hat{D}_\varphi H_N^s. \tag{7.41}$$

It is always the case that

$$|\varepsilon_1\rangle \in \ker\{D_\varphi^s\}, \tag{7.42}$$

which implies

$$\dim \ker\{D_\varphi^s\} \geq 1. \tag{7.43}$$

This inequality can easily be tested using the singular value decomposition of $D_\varphi^s$: at least one of its singular values must be zero. Eigenvalues $\varphi_k$ come in conjugate pairs where one of each pair has its argument in $[0, \pi[$. As a consequence, we can decompose $D_\theta^s$ for multiple values of $\theta$ between 0 and $\pi$ until we find all the eigenvalues of $Q$ associated with eigenvectors in $\mathcal{E}$, provided that $D_\theta^s$ computation is not too complex. The dimension of the kernel gives us the multiplicity of the eigenvalue $e^{i\varphi_k}$. It is possible to halve the duration of the eigenvalue search by noticing that

$$D_\varphi^s = -D_{-\varphi}^s, \tag{7.44}$$

and that

$$\ker\{D_\varphi^s\} = \ker\{D_{-\varphi}^s\}. \tag{7.45}$$

Therefore, it is possible to restrain the search to the $[0, \pi/2]$ span.

### 7.2 Efficient computation of the criterion

The direct computation of $D_\theta^s$ becomes impossible on a classical computer if $n$ is big, as it involves a $2^n \times 2^n$ matrix. However, we have

$$D_\theta^s = \sum_{w_p=0}^n \hat{D}_\theta(w_p) H_N^{s,w_p\,\mathrm{T}} H_N^{s,w_p}. \tag{7.46}$$

Let us define $\Xi_{w_p}$ as the matrix product $H_N^{s,w_p\,\mathrm{T}} H_N^{s,w_p}$. We have

$$D_\theta^s = \sum_{w_p=0}^n \hat{D}_\theta(w_p) \Xi_{w_p}. \tag{7.47}$$

We will see that $\Xi_{w_p}$ can be efficiently computed without the need of $H_N^s$. Let us note $a$ the position of a given solution, $\alpha$ its binary representation, and $|h_a\rangle$ the corresponding column in $\bar{H}_N^{w_p}$, the unnormalized Hadamard matrix restricted to the rows whose indices have a Hamming weight of $w_p$ defined in Sect. 1.7. The elements of $\Xi_w$ are

$$\Xi_w(a, b) = \frac{1}{N} \langle h_a | h_b \rangle, \tag{7.48}$$

$$= \frac{1}{N} \sum_p h_a(p) h_b(p), \tag{7.49}$$

The unnormalized Hadamard matrix has an interesting property about its values:

$$\bar{H}_N(a,b) = (-1)^{\langle \alpha | \beta \rangle}, \tag{7.50}$$

where $\alpha$ and $\beta$ are the binary representation of $a$ and $b$. Therefore,

$$h_a(p) h_b(p) = (-1)^{\langle \rho | \alpha \rangle} (-1)^{\langle \rho | \beta \rangle}, \tag{7.51}$$
$$= (-1)^{\langle \rho | \alpha \oplus \beta \rangle}, \tag{7.52}$$
$$= h_{a \oplus b}(p), \tag{7.53}$$

where $a \oplus b$ is the position whose binary index is $\alpha \oplus \beta$. Thus,

$$\Xi_w(a,b) = \frac{1}{N} \sum_p h_{a \oplus b}(p). \tag{7.54}$$

Denote by $\eta_m(w_p)$ the sum of all elements of $h_m$ whose weight is $w_p$.

$$\eta_m(w_p) = \sum_{w_p} h_m(p), \tag{7.55}$$
$$= \sum_{w_p} (-1)^{\langle \rho | \mu \rangle}. \tag{7.56}$$

The elements of the sum are $+1$ when $\langle \rho | \mu \rangle$ is even, which happens when $\rho$ contains an even number of 1 inside the $w_m$ positions corresponding to 1 in $\mu$ (the Hamming weight of $\mu$ is $w_m$). Denote this number by $2l$. There are $\binom{w_m}{2l}$ possible placements for these 1, and $\binom{n-w_m}{w_p-2l}$ for the $w_p - 2l$ remaining 1. The total number of $+1$ in the previous sum is then

$$\zeta_m(w_p) = \sum_l \binom{w_m}{2l} \binom{n - w_m}{w_p - 2l}, \tag{7.57}$$

for all values of $l$ for which the binomial coefficient exists, that is

$$0 \leq l \leq \frac{m}{2} \quad \text{and} \tag{7.58}$$
$$\frac{m+k-n}{2} \leq l \leq \frac{k}{2}. \tag{7.59}$$

Since the number of indices associated with $w_p$ is $\binom{n}{w_p}$, we have $\binom{n}{w_p} - \zeta_m(w_p)$ times $-1$ in the sum and

$$\eta_m(w_p) = \zeta_m(w_p) - \left( \binom{n}{w_p} - \zeta_m(w_p) \right), \tag{7.60}$$

so
$$\eta_m(w_p) = 2\zeta_m(w_p) - \binom{n}{w_p}. \tag{7.61}$$

If $m = 0$, we have $\binom{0}{0} = 1$ and $\eta_0(w_p) = \binom{n}{w_p}$.

Note that $\eta_m(w_p)$ does not depend on $m$ but on $w_m$, which allows us to compute it for several solutions at the same time. Therefore, we can denote it by $\eta(w_p, w_m)$. Finally

$$\Xi_{w_p}(a, b) = \frac{1}{N}\eta(w_p, w_{a\oplus b}). \tag{7.62}$$

Thanks to this result, it is possible to compute $D_\theta^s$, and therefore the criterion, in a polynomial time with a classical computer.

### 7.3 Components of the key vectors in the eigenspace of interest

#### 7.3.1 Regular cases

In order to compute the probability of success, we need to compute the components of $|u\rangle$ and $|s\rangle$ in $\mathcal{E}$. Once the eigenvalues $\lambda_k = e^{i\varphi_k}$ are determined, we can compute the vectors $|\varepsilon_1\rangle$ in the eigenspace associated with each $\lambda_k$ with Equation (7.42). For a given eigenvalue $\lambda = e^{i\varphi}$

$$|\hat{e}_-\rangle = H_N^s|\varepsilon_1\rangle, \tag{7.63}$$
$$|\hat{e}_+\rangle = i\hat{D}_\varphi|\hat{e}_-\rangle, \tag{7.64}$$
$$|\hat{\varepsilon}_+\rangle = |\hat{e}_+\rangle \otimes |u_n\rangle, \tag{7.65}$$
$$|\hat{\varepsilon}_-\rangle = -i\hat{\Gamma}_\varphi^{-1}|\hat{\varepsilon}_+\rangle, \tag{7.66}$$
$$|\hat{\varepsilon}\rangle = |\hat{\varepsilon}_+\rangle + |\hat{\varepsilon}_-\rangle, \tag{7.67}$$
$$|\varepsilon\rangle = F|\hat{\varepsilon}\rangle. \tag{7.68}$$

We also have
$$|\varepsilon\rangle = L_1|\varepsilon_1\rangle + L_2|\varepsilon_2\rangle + FL_3^{\prime o}|\varepsilon_3\rangle, \tag{7.69}$$

so
$$\langle\varepsilon|s\rangle = \langle\varepsilon_1|L_1^T|s\rangle, \tag{7.70}$$
$$= \langle\varepsilon_1|u_M\rangle, \tag{7.71}$$

and
$$\langle\varepsilon|u\rangle = \langle\varepsilon_1|L_1^T|u\rangle + \langle\varepsilon_2|L_2^T|u\rangle, \tag{7.72}$$
$$= \sqrt{\frac{M}{N}}\langle\varepsilon_1|u_M\rangle + \langle\varepsilon_+|u\rangle, \tag{7.73}$$

where

$$\langle \varepsilon_+ | u \rangle = \langle e_+ \otimes u_n | u_N \otimes u_n \rangle, \quad (7.74)$$
$$= \langle e_+ | u_N \rangle. \quad (7.75)$$

Recall that $\langle e_+ | u_N \rangle$ is equal to the average value of $\langle e_+ |$ and that the first term of a Fourier transform is the average value of the transformed vector, therefore,

$$\langle \varepsilon_+ | u \rangle = \langle \hat{e}_+(0) |, \quad (7.76)$$
$$= -i \hat{D}_\varphi(0) \langle \hat{e}_-(0) |, \quad (7.77)$$
$$= -i \cot \frac{\varphi}{2} \langle e_- | u_n \rangle, \quad (7.78)$$
$$= -i \cot \frac{\varphi}{2} \sqrt{\frac{M}{N}} \langle \varepsilon_1 | u_M \rangle, \quad (7.79)$$

and then

$$\langle \varepsilon | u \rangle = \sqrt{\frac{M}{N}} \left( 1 - i \cot \frac{\varphi}{2} \right) \langle \varepsilon | s \rangle. \quad (7.80)$$

We now have the components of $|s\rangle$ and $|u\rangle$ in $\mathcal{E}$ for each $|\varepsilon_1\rangle$ associated with each eigenvalue $\lambda_k$:

$$s'(k, l) = \langle \varepsilon_1 | u_M \rangle, \quad (7.81)$$
$$u'(k, l) = \sqrt{\frac{M}{N}} \left( 1 - i \cot \frac{\varphi_k}{2} \right) s'(k, l). \quad (7.82)$$

Note that we use the notations $s'(k, l)$ and $u'(k, l)$ instead of $s(k, l)$ and $u(k, l)$. This is because the vectors $|\varepsilon\rangle$ are not necessarily orthogonal nor unit vectors, so these equations do not directly provide the projections of $|u\rangle$ and $|s\rangle$ in the eigenspace associated with $e^{i\varphi}$. A correction is required. First, we define the transformation $f$, identical to the one seen in Sect. 7.2 by

$$f(A) = \sum_{w_p} A(w_p) H_N^{s,w_p \mathrm{T}} H_N^{s,w_p}, \quad (7.83)$$

where $A(w_p)$ is a diagonal matrix whose elements are functions of $w_p$ only.

Due to the orthogonality of $|\hat{\varepsilon}_-\rangle$ and $|\hat{\varepsilon}_+\rangle$, for two given vectors $|\hat{\varepsilon}\rangle$ and $|\hat{\varepsilon}'\rangle$, we have

$$\langle \hat{\varepsilon} | \hat{\varepsilon}' \rangle = \langle \hat{\varepsilon}_- | \hat{\varepsilon}'_- \rangle + \langle \hat{\varepsilon}_+ | \hat{\varepsilon}'_+ \rangle, \quad (7.84)$$

and so, for each $|\varepsilon_1\rangle$, we have

$$\langle \hat{\varepsilon}_+ | \hat{\varepsilon}'_+ \rangle = \langle \hat{e}_+ \otimes u_n | \hat{e}'_+ \otimes u_n \rangle, \quad (7.85)$$
$$= \langle e_+ | e'_+ \rangle, \quad (7.86)$$
$$= \langle e_- | \hat{D}_\varphi^2 | e'_- \rangle, \quad (7.87)$$

$$\langle \varepsilon_1 | H_N^{s\,\mathrm{T}} \hat{D}_\varphi^2 H_N^s | \varepsilon_1' \rangle, \tag{7.88}$$

$$= \langle \varepsilon_1 | f\left( \hat{D}_\varphi^2 \right) | \varepsilon_1' \rangle, \tag{7.89}$$

and

$$\langle \hat{\varepsilon}_- | \hat{\varepsilon}_-' \rangle = \langle \hat{\varepsilon}_+ | \hat{\Gamma}_\varphi^{-2} | \hat{\varepsilon}_+' \rangle, \tag{7.90}$$

$$= \langle e_+ \otimes u_n | \hat{\Gamma}_\varphi^{-2} | e_+' \otimes u_n \rangle, \tag{7.91}$$

$$= \langle e_+ | T_\varphi | e_+' \rangle, \tag{7.92}$$

$$= \langle e_- | \hat{D}_\varphi T_\varphi \hat{D}_\varphi | e_-' \rangle, \tag{7.93}$$

$$= \langle \varepsilon_1 | H_N^{s\,\mathrm{T}} \hat{D}_\varphi T_\varphi \hat{D}_\varphi H_N^s | \varepsilon_1' \rangle, \tag{7.94}$$

$$= \langle \varepsilon_1 | f\left( \hat{D}_\varphi^2 T_\varphi \right) | \varepsilon_1' \rangle, \tag{7.95}$$

where $T_\varphi$ is a diagonal matrix such that the $T_\varphi(p)$ contains the average value of the diagonal of $\hat{\Gamma}_\varphi^{-2}$ over block $p$, that is

$$T_\varphi(p) = \left(1 - \frac{w_p}{n}\right) \tan^2 \frac{\varphi}{2} + \frac{w_p}{n} \cot^2 \frac{\varphi}{2}. \tag{7.96}$$

Finally, we have

$$\langle \hat{\varepsilon} | \hat{\varepsilon}' \rangle = \langle \varepsilon_1 | f\left( \hat{D}_\varphi^2 + \hat{D}_\varphi^2 T_\varphi \right) | \varepsilon_1' \rangle. \tag{7.97}$$

Let $E_1$ be a matrix whose columns are the $|\varepsilon_1\rangle$ in the eigenspace associated with $\lambda_k$. Then,

$$E^\dagger E = E_1^\dagger f\left( \hat{D}_\varphi^2 + \hat{D}_\varphi^2 T_\varphi \right) E_1, \tag{7.98}$$

where each column of $E$ is an eigenvector $|\hat{\varepsilon}\rangle$ in the same eigenspace and $E^\dagger E$ is a correlation matrix from which we will deduce the correction. We can diagonalize $E^\dagger E$ via

$$E^\dagger E = V_E S_E^2 V_E^\dagger, \tag{7.99}$$

where $S_E^2$ is a diagonal matrix containing the eigenvalues of $E^\dagger E$ and $V_E$ a matrix whose columns are its eigenvectors. We can deduce the singular value decomposition of $E$. Since

$$E = U_E S_E V_E^\dagger, \tag{7.100}$$

where $U_E$ is an $N \times M$ matrix and $S_E$ and $V_E$ are $M \times M$ matrices. Therefore

$$s'(k, l) = \langle E(l) | s \rangle, \tag{7.101}$$

where $|E(l)\rangle$ is the $l$-th column of $E$. Because the columns of $U_E$ are orthogonal unit vectors, we can use it to compute the correct projection of $|s\rangle$ in $\mathcal{E}$, that is

$$|s(k)\rangle = U_E | s \rangle. \tag{7.102}$$

Note that $U_E$ has $N_e$ rows and as many columns as there are eigenvectors associated to $\lambda$, so it may be too big for a classical computer to handle. However, its computation can be avoided since

$$|s(k)\rangle = S_E^{-1} V_E^{-1} |s'(k)\rangle, \qquad (7.103)$$

where $S_E$ and $V_E$ are square matrices whose size is dim ker $\{D_\varphi^s\}$.

Finally

$$u(k,l) = \sqrt{\frac{M}{N}} \left(1 - i \cot \frac{\varphi_k}{2}\right) s(k,l). \qquad (7.104)$$

### 7.3.2 Real eigenvalues case

We will now consider the case $\lambda = +1$. This eigenvalue implies that $P_+^S|\varepsilon_-\rangle = 0$ and $P_-^S|\varepsilon_+\rangle = 0$. Then, $|\varepsilon_-\rangle \in E_{-,-}^{S,CO}$ and $|\varepsilon_+\rangle \in E_{+,+}^{S,CO}$, but we have shown that dim $E_{+,+}^{S,CO} = 0$, so $|\varepsilon\rangle \in E_{-,-}^{S,CO}$. The intersection of $E_{-,-}^{S,CO}$ with $\mathcal{E}$ has a dimension of $M - 1$. Similarly, if $\lambda = -1$, we have $P_-^S|\varepsilon_-\rangle = 0$ and $P_+^S|\varepsilon_+\rangle = 0$. Then, $|\varepsilon_-\rangle \in E_{+,-}^{S,CO}$ and $|\varepsilon_+\rangle \in E_{-,+}^{S,CO}$, but dim $E_{-,+}^{S,CO} = 0$, so $|\varepsilon\rangle \in E_{+,-}^{S,CO}$, whose dimension in the complement is also $M - 1$. In the following, we will only consider $\lambda = -1$, as the discussion related to $\lambda = +1$ is similar, but of less interest since both $|u\rangle$ and $|s\rangle$ have null projections over this eigenspace. We have shown in Sect. 4 that the subspace of $E_{-,-}^{S,CO}$ in the complement of $E_{-,-}^{S,C}$, that is in $\mathcal{E}$, has a dimension of $M - 1$. Furthermore, we established a parametric form of the corresponding eigenvectors:

$$|\varepsilon\rangle = FL_1|\varepsilon_1\rangle + F_3'^\circ|\varepsilon_3\rangle, \qquad (7.105)$$

with a constraint on $|\varepsilon_1\rangle$ that is $\langle\varepsilon_1|h\rangle = 0$, $\langle h|$ being the last row of $H_N^s$. It also follows that

$$|\varepsilon_3(p)\rangle = -\sqrt{\frac{w_p}{n - w_p}} \left(H_N^s|\varepsilon_1\rangle\right)_p. \qquad (7.106)$$

Since $|u\rangle$ and $|s\rangle$ are both orthogonal to $L_3'^\circ$, we have

$$\langle\varepsilon|s\rangle = \langle\varepsilon_1|L_1^\dagger|s\rangle, \qquad (7.107)$$
$$= \langle\varepsilon_1|u_M\rangle, \qquad (7.108)$$

and

$$\langle\varepsilon|u\rangle = \langle\varepsilon_1|L_1^\dagger|u\rangle, \qquad (7.109)$$
$$= \sqrt{\frac{M}{N}} \langle\varepsilon_1|u_M\rangle, \qquad (7.110)$$
$$= \sqrt{\frac{M}{N}} \langle\varepsilon|s\rangle. \qquad (7.111)$$

As for the previous case, the generated $|\varepsilon\rangle$ are not orthogonal and must be corrected in a similar way.

Due to the orthogonality of $L_1$ and $L_3'^o$, we have

$$\langle \varepsilon | \varepsilon' \rangle = \langle \varepsilon_1 | \varepsilon_1' \rangle + \langle \varepsilon_3 | \varepsilon_3' \rangle. \tag{7.112}$$

Let $W$ be the diagonal matrix whose elements are $W(p) = w_p/(n - w_p)$. Then

$$\langle \varepsilon_3 | \varepsilon_3' \rangle = \langle \varepsilon_1 | H_N^{s\,\mathrm{T}} W H_N^s | \varepsilon_1' \rangle, \tag{7.113}$$
$$= \langle \varepsilon_1 | f(W) | \varepsilon_1' \rangle. \tag{7.114}$$

The corresponding correlation matrix can be computed via

$$E^\dagger E = E_1^\dagger f(I_N + W) E_1. \tag{7.115}$$

From this point, the correction method is the same as in the previous case.

### 7.3.3 Singular $\hat{D}_\varphi^{-1}$ case

The last case to cover is the one where $\hat{D}_\varphi^{-1}$ is not invertible because it contains at least one zero value on its diagonal. According to Equation (7.33), this happens when

$$\left(1 - \frac{w_p}{n}\right) \tan \frac{\varphi}{2} - \frac{w_p}{2} \cot \frac{\varphi}{2} = 0, \tag{7.116}$$

that is

$$\left(1 - \frac{w_p}{n}\right) \tan^2 \frac{\varphi}{2} = \frac{w_p}{2}, \tag{7.117}$$
$$\left(1 - \frac{w_p}{n}\right) \frac{1 - \cos\varphi}{1 + \cos\varphi} = \frac{w_p}{2}, \tag{7.118}$$

so

$$\cos\varphi = 1 - \frac{2 w_p}{n}, \tag{7.119}$$

which corresponds to the complex eigenvalues of $U$, $\lambda = \lambda_{w_p}$. In this case, $\hat{D}_\varphi^{-1}(p) = 0$, and Eq. (7.36) implies that $|\hat{e}_-(p)\rangle = 0$. Since

$$|\hat{e}_-\rangle = H_N^s |\varepsilon_1\rangle, \tag{7.120}$$

we have a new constraint on $|\varepsilon_1\rangle$, so

$$|\varepsilon_1\rangle \in \ker\left\{ H_N^{s,w_p} \right\}. \tag{7.121}$$

As $H_N^{s,w_p}$ is a large matrix, its kernel can be complex to compute, but since for any operator

$$\ker\{A\} = \ker\left\{ A^\dagger A \right\}, \tag{7.122}$$

we have
$$|\varepsilon_1\rangle \in \ker\{\Xi_{w_p}\}, \qquad (7.123)$$

where $\Xi_{w_p} = H_N^{s,w_p\,\mathrm{T}} H_N^{s,w_p}$. We showed in Sect. 7.2 that $\Xi_{w_p}$ can be computed easily.

Equation (7.37) is only valid over the positions $\bar{p}$ whose binary representation Hamming weight is not $w_p$. We have

$$|\hat{e}_+(\bar{p})\rangle = i\hat{D}_\varphi(\bar{p})|\hat{e}_-(\bar{p})\rangle. \qquad (7.124)$$

We define $|\hat{e}_+^{\bar{p}}\rangle$ as the vector whose elements at positions $p$ are zero and others are determined by the equation above. It can be obtained by using a modified $\hat{D}_\varphi$ whose elements at positions $p$ are zero. Let us denote this modified matrix $\hat{D}_\varphi^{\bar{p}}$. Let us note $|x\rangle$ the length $\binom{n}{w_p}$ vector containing the elements of $|\hat{e}_+\rangle$ at positions $p$. We have

$$|\hat{e}_+\rangle = |\hat{e}_+^{\bar{p}}\rangle + I_N^{w_p}|x\rangle. \qquad (7.125)$$

Since $|\hat{e}_+\rangle \in \mathrm{span}\{H_N^{\bar{s}}\}$, its $M$ components corresponding to solutions must be zero. Therefore,
$$H_N^{s\,\mathrm{T}}|\hat{e}_+\rangle = 0, \qquad (7.126)$$

and
$$iH_N^{s\,\mathrm{T}}\hat{D}_\varphi^{\bar{p}}|\hat{e}_-\rangle + H_N^{s,w_p\,\mathrm{T}}|x\rangle = 0, \qquad (7.127)$$

which we denote
$$iD_\varphi^{s,\bar{p}}|\varepsilon_1\rangle + H_N^{s,w_p\,\mathrm{T}}|x\rangle = 0. \qquad (7.128)$$

We have
$$\begin{bmatrix}|\varepsilon_1\rangle \\ |x\rangle\end{bmatrix} \in \ker\left\{\begin{bmatrix}iD_\varphi^{s,\bar{p}} & H_N^{s,w_p\,\mathrm{T}}\end{bmatrix}\right\}. \qquad (7.129)$$

In order to deal with this constraint and the one from Eq. (7.123), we introduce a matrix $Y$ whose columns are an orthonormal basis of $\ker\{\Xi_{w_p}\}$. We have
$$|\varepsilon_1\rangle = Y|\hat{\varepsilon}_1\rangle. \qquad (7.130)$$

The matrix $H_N^{s,w_p\,\mathrm{T}}$ can be huge, so computing its kernel may be difficult. However, this can be avoided using the properties of the singular value decomposition. Let the SVD of $H_N^{s,w_p}$ be
$$H_N^{s,w_p} = U_{w_p} S_{w_p} V_{w_p}^\dagger. \qquad (7.131)$$

We have
$$H_N^{s,w_p\,\mathrm{T}} = V_{w_p} S_{w_p} U_{w_p}^\dagger. \qquad (7.132)$$

Any vector $|a\rangle$ in $\mathrm{span}\{H_N^{s,w_p\,\mathrm{T}}\}$ can be written:
$$|a\rangle = H_N^{s,w_p\,\mathrm{T}}|b\rangle, \qquad (7.133)$$

$$= V_{w_p} S_{w_p} U_{w_p}^\dagger |b\rangle, \tag{7.134}$$

$$= V_{w_p} S_{w_p} |b_\parallel\rangle, \tag{7.135}$$

where $|b_\parallel\rangle = U_{w_p}^\dagger |b\rangle$. Conversely

$$|b\rangle = U_{w_p} |b_\parallel\rangle + |b_\perp\rangle, \tag{7.136}$$

where $|b_\perp\rangle$ is the component of $|b\rangle$ orthogonal to span $\{U_{w_p}\}$. Since $|b_\perp\rangle$ has no impact on $|a\rangle$, it can be considered null. Then,

$$|b\rangle = U_{w_p} |b_\parallel\rangle, \tag{7.137}$$

and, therefore,

$$\text{span}\left\{H_N^{s,w_p\,\text{T}}\right\} = \text{span}\left\{V_{w_p} S_{w_p}\right\}. \tag{7.138}$$

Also, note that for two vectors $|b\rangle$ and $|b'\rangle$

$$\langle b|b'\rangle = \left(\langle b_\parallel | U_{w_p}^\dagger + \langle b_\perp|\right)\left(U_{w_p}|b'_\parallel\rangle + |b'_\perp\rangle\right), \tag{7.139}$$

$$= \left\langle b_\parallel \left| U^\dagger U \right| \tilde{b}'\right\rangle + \langle b_\perp|b'_\perp\rangle, \tag{7.140}$$

$$= \langle b_\parallel|b'_\parallel\rangle + \langle b_\perp|b'_\perp\rangle, \tag{7.141}$$

$$= \langle b_\parallel|b'_\parallel\rangle, \tag{7.142}$$

since $|b_\perp\rangle$ can be forced to zero.

$V_{w_p}$ and $S_{w_p}$ can easily be computed from the SVD of $\Xi_{w_p}$, an $M \times M$ matrix. Then,

$$\Xi_{w_p} = V_{w_p} S_{w_p}^2 V_{w_p}^\dagger. \tag{7.143}$$

Then, Eq. (7.129) becomes

$$\begin{bmatrix} |\hat{\varepsilon}_1\rangle \\ |\tilde{x}\rangle \end{bmatrix} \in \ker\left\{\begin{bmatrix} iD_\varphi^{s,\bar{p}} Y & V_{w_p} S_{w_p} \end{bmatrix}\right\}, \tag{7.144}$$

where $|\tilde{x}\rangle = U_{w_p}^\dagger |x\rangle$ is a length rank $\{\Xi_{w_p}\}$ vector. We now have

$$\langle \hat{\varepsilon}_+|\hat{\varepsilon}'_+\rangle = \langle \hat{e}_+|\hat{e}'_+\rangle, \tag{7.145}$$

$$= \langle \hat{e}_+^{\bar{p}}|\hat{e}'^{\bar{p}}_+\rangle + \langle \tilde{x}|\tilde{x}'\rangle, \tag{7.146}$$

$$= \langle \hat{e}_-|\left(\hat{D}_\varphi^{s,\bar{p}}\right)^2|\hat{e}'_-\rangle + \langle \tilde{x}|\tilde{x}'\rangle, \tag{7.147}$$

$$= \langle \varepsilon_1|f\left(\left(\hat{D}_\varphi^{s,\bar{p}}\right)^2\right)|\varepsilon'_1\rangle + \langle \tilde{x}|\tilde{x}'\rangle, \tag{7.148}$$

and

$$\langle \hat{\varepsilon}_-|\hat{\varepsilon}'_-\rangle = \langle \hat{\varepsilon}_+|\hat{\Gamma}_\varphi^{-2}|\hat{\varepsilon}'_+\rangle, \quad (7.149)$$

$$= \langle \hat{e}_+ \otimes u_n|\hat{\Gamma}_\varphi^{-2}|\hat{e}'_+ \otimes u_n\rangle, \quad (7.150)$$

$$= \langle \hat{e}_+|W_\varphi|\hat{e}'_+\rangle + \langle \tilde{x}|I_N^{w_p\mathrm{T}} W_\varphi I_N^{w_p}|\tilde{x}'\rangle. \quad (7.151)$$

Since $\cos\varphi = 1 - 2w_p/n$,

$$\tan\frac{\varphi}{2} = \sqrt{\frac{w_p}{n-w_p}}, \quad (7.152)$$

$$\cot\frac{\varphi}{2} = \sqrt{\frac{n-w_p}{w_p}}. \quad (7.153)$$

Then, at positions $p$

$$W_\varphi(p) = \frac{n-w_p}{n}\tan^2\frac{\varphi}{2} + \frac{w_p}{n}\cot^2\frac{\varphi}{2}, \quad (7.154)$$

$$= \frac{w_p}{n} + \frac{n-w_p}{n}, \quad (7.155)$$

$$= 1, \quad (7.156)$$

and we have

$$\langle \hat{\varepsilon}_-|\hat{\varepsilon}'_-\rangle = \langle \hat{e}_-|\hat{D}_\varphi^{\bar{p}} W_\varphi \hat{D}_\varphi^{\bar{p}}|\hat{e}'_-\rangle + \langle \tilde{x}|\tilde{x}'\rangle, \quad (7.157)$$

$$= \langle \varepsilon_1|f\left(\left(\hat{D}_\varphi^{\bar{p}}\right)^2 W_\varphi\right)|\varepsilon'_1\rangle + \langle \tilde{x}|\tilde{x}'\rangle. \quad (7.158)$$

If we denote by $X$ the matrix whose columns are the vectors $|\tilde{x}\rangle$, the correlation matrix $E^\dagger E$ is given by

$$E^\dagger E = E_1^\dagger f\left(\left(\hat{D}_\varphi^{\bar{p}}\right)^2 + \left(\hat{D}_\varphi^{\bar{p}}\right)^2 W_\varphi\right) E_1 + 2X^\dagger X. \quad (7.159)$$

The computation of the components of $|u\rangle$ and $|s\rangle$ in $\mathcal{E}$ is done in the same way as the previous cases: The computation of $\langle \varepsilon|s\rangle$ only involves $L_1$, which is orthogonal to $|x\rangle$, and the computation of $\langle \varepsilon|u\rangle$ involves $\langle \varepsilon_+|u\rangle$ which depends on $L_2$. This could lead to the appearance of $|x\rangle$. However, we see that $\langle \varepsilon_+|u\rangle = |\hat{e}_+(0)\rangle$, which is never affected by $|x\rangle$. Indeed, $|x\rangle$ only affects the positions whose weight is $w_p$, and in the case we are currently studying, $w_p \in [1, n-1]$. Therefore, $|x\rangle$ does not appear in the computation of $\langle \varepsilon|u\rangle$.

# 8 Simulation and results

## 8.1 Procedure

As a consequence of the results obtained in Sect. 7, it is possible to design a simple procedure to compute $p_t$ as a function of $t$:

1. Compute the dimension of $\mathcal{E}$ according to Eq. 6.36 and create with an arbitrary chosen step a linearly spaced set of angle $\theta$ from 0 to $\pi/2$.
2. Compute the $D_\theta^s$ according to the method seen in Sect. 7.2 and their least singular values.
3. Find all local minima of the criterion, which is the lowest singular value of each $D_\theta^s$ as a function of $\theta$. Those local minima are found where $\theta$ equals one of the $\varphi$.
4. Compute the $S$ and $V$ matrices from the SVD of each $D_\varphi^s$. The number of singular values equal to zero is the multiplicity $m$ of the eigenvalue $e^{i\varphi}$. Discard eigenvalues with a multiplicity of $m = 0$. Add the symmetric eigenvalues from the $]\pi/2, \pi]$ span to the found eigenvalues.
5. For each $\varphi$, extract the $m$ last columns of $V$ as the vectors $|\varepsilon_1\rangle$. Then for each vector $|\varepsilon_1\rangle$ of each eigenvalue, compute $s'(k, l)$ according to 7.81. Correct the $|s(k)\rangle$ according to Eq. 7.103 and compute the $|u(k)\rangle$ according to Eq. 7.104. Then, compute the $|u\rangle$ and $|s\rangle$ components associated with the conjugate eigenvalues, that is the conjugates of all $u(k, l)$ and $s(k, l)$ computed above.
6. Compute the $M - 1$ $|\varepsilon_1\rangle$ vectors associated with the $-1$ eigenvalue as the kernel of $\langle h|$, the last row of $H_N^s$. Compute the $s'(k, l)$ according to Eq. 7.108 and correct the $|s(k)\rangle$ in the same manner as before with the correlation matrix obtained in Eq. 7.115. Deduce the corresponding $|u(k)\rangle$. If the total number of $s(k, l)$ found is equal to $\dim \mathcal{E}$, go to step 9.
7. Compute all $\varphi$ that respect Eq. 7.119 and their $D_\varphi^{\bar{p}}$ as described in Sect. 7.3.3. Compute the SVD of the $D_\varphi^{s, \bar{p}}$ and keep those whose least singular value is 0. Then, compute the $s'(k, l)$ in the same manner as before, using $D_\varphi^{s, \bar{p}}$ instead of $D_\varphi^s$. Correct the $|s(k)\rangle$ in the same manner as before with the correlation matrix obtained in Eq. 7.159. Finally, compute the corresponding $|u(k)\rangle$.
8. If the number of $s(k, l)$ found is still lesser than $\dim \mathcal{E}$, one can choose to start the procedure again with a smaller step on $\theta$ or continue with a less precise simulation.
9. Compute the probability of success $p_t$ for all values of $t$ between 0 and a chosen number of iterations according to Eq. 7.3.

## 8.2 Example

In this section, we will show an overview of the proposed procedure for a 6-dimension hypercube. The results are summarized in Figs. 12 and 13. Note that here, the eigenvalues search is done in the $[0, \pi]$ span. The results plotted in Fig. 13 are compared to those obtained by direct simulation using the walk operator $Q$. The probability of success obtained by this method is computed according to Eq. 0.2, implying the computation of $Q^t$, which becomes exponentially long as $n$ increases. The results of both simulation methods match closely as predicted.

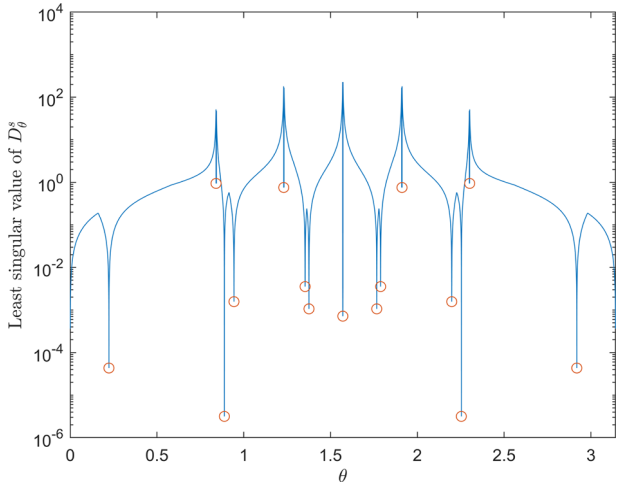

**Fig. 12** Search for the eigenvalues phases with $n = 6$, $M = 2$ solutions at position 3 and 6 and a step $\Delta\theta = \pi/10\,000$. Found eigenvalues are circled

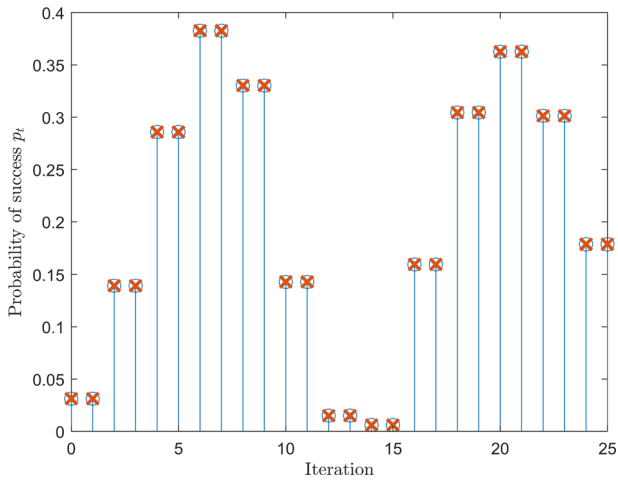

**Fig. 13** Evolution of the probability of success in function of the number of iterations. The circles correspond to our method results while the crosses are the exact values obtained by direct simulation. Simulation with $n = 6$, $M = 2$ solutions at position 3 and 6

First, we compute $\dim \mathcal{E}$. It is not required for the simulation itself, but it allows us to check if we find enough eigenvalues. In this example, we find $\dim \mathcal{E} = 22$.

We chose a step of $\Delta\theta = \pi/10\,000$, but $\pi/500$ would be precise enough to obtain a good approximation. Over the 10 000 angle values tested, we find 15 local minima in the $[0, \pi]$ span. After computing the SVD of the corresponding $D_\theta^s$ matrices, we can find that 4 of them do not have any singular value equal to zero. These correspond to the 4 highest local minima seen in Fig. 12. Additionally, the $D_\theta^s$ matrix for $\theta = \pi/2$ is undefined, as $\hat{D}_{\pi/2}^{-1}$ is singular. In this case, we apply Sect. 7.3.3.

**Table 4** Estimated nonzero components of $|s\rangle$ and $|u\rangle$ in $\mathcal{E}$ with $n = 6$, $M = 2$ solutions at position 3 and 6

| $\varphi$ | $|s(k)\rangle$ | $|u(k)\rangle$ |
|---|---|---|
| 0.2231 | 0.4382 | 0.0775 − 0.6917i |
| 0.9434 | 0.1769 | 0.0313 − 0.0613i |
| 1.3755 | −0.1632 | −0.0289 + 0.0351i |
| 1.7661 | 0.1632 | 0.0289 − 0.0237i |
| 2.1982 | −0.1769 | −0.0313 + 0.0160i |
| 2.9185 | −0.4382 | −0.0775 + 0.0087i |
| −2.9185 | −0.4382 | −0.0775 − 0.0087i |
| −2.1982 | −0.1769 | −0.0313 − 0.0160i |
| −1.7661 | 0.1632 | 0.0289 + 0.0237i |
| −1.3755 | −0.1632 | −0.0289 − 0.0351i |
| −0.9434 | 0.1769 | 0.0313 + 0.0613i |
| −0.2231 | 0.4382 | 0.0775 + 0.6917i |

Then, we add the symmetric eigenvalues over $]\pi, 2\pi[$ and the eigenvalue $-1$ with a multiplicity of $M - 1 = 1$.

As in any numerical computation, deciding under which threshold a very small singular value is considered zero is always somewhat arbitrary. In some cases, depending on this threshold, the procedure can produce more eigenvalues than expected. These excess values are not an issue, as they correspond to vectors outside $\mathcal{E}$. Therefore, they give zero components to $|s\rangle$ and $|u\rangle$ and do not affect the computation.

Then, we can compute the corresponding components of $|s\rangle$ and $|u\rangle$ in $\mathcal{E}$. Note that in this example, among the 22 components of $|s\rangle$ and $|u\rangle$, only 12 are nonzero. These are displayed in Table 4.

### 8.3 Upper bound for the probability of success

From the values of $\varphi_k$ and $s(k, l)$, it is possible to compute an upper bound for the probability of success. Indeed, we have

$$p_t = \left| \sum_{k,l} s(k,l)^* e^{i\varphi_k t} u(k,l) \right|^2. \tag{8.1}$$

Using the triangle inequality, we obtain

$$p_t \leq \left( \sum_{k,l} |s(k,l)^* e^{i\varphi_k t} u(k,l)| \right)^2, \tag{8.2}$$

$$\leq \left( \sum_{k,l} |s(k,l)| |u(k,l)| \right)^2. \tag{8.3}$$

From Eq. 7.104, we have

$$|u(k,l)| = \sqrt{\frac{M}{N}} \sqrt{1 + \cot^2 \frac{\varphi_k}{2}} |s(k,l)|, \tag{8.4}$$

and finally

$$p_t \leq \frac{M}{N} \left( \sum_k \sqrt{1 + \cot^2 \frac{\varphi_k}{2}} \sum_l |s(k,l)|^2 \right)^2. \tag{8.5}$$

In the example shown in Sect. 8.2, we obtain $p_t \leq 0.5509$. We checked the validity of this upper bound: after 10 000 iterations, the maximum value reached by the probability of success is 0.4279, which is, as expected, under the upper bound (here it is 28% under the bound). While the upper bound is not a close one in this example, such a theoretical bound is always interesting to have a fast estimate of the best possible performance that the quantum walk search could reach.

## Conclusion

Quantum random walk search is a promising algorithm. However its theoretical study can be extremely complex and possibly intractable, especially when there is more than one solution. In this paper, we have proposed a method which allows us to compute the probability of success of the algorithm as a function of the number of iterations, without simulating the search itself. It is therefore possible to compute that probability on a classical computer even for search whose state space dimension is very large.

Knowledge of the probability of success allows us to determine the optimal time of measurement (the time which maximizes this probability), provided that we know the number of solutions and their relative positions. For instance, we may know that the acceptable solutions are the interior of a hypersphere of a given radius, whose absolute position is unknown.

It is also a powerful tool for in-depth study and better understanding of quantum random walk search properties. For instance, in our future work we plan to study how the number of solutions and their relative positions impact the probability of success. This may highlight a phenomenon of interference between solutions which does not exist in a classical search.

The study of the subspace $\mathcal{E}$ could also be transposed over different walks patterns, such as planar quantum walks, possibly leading to interesting results. A further study could be considered in order to find the exact complexity of the method, or the required precision on $\theta$ during the eigenvalues search.

**Acknowledgements** The authors thank the Brest Institute of Computer Science and Mathematics (IBNM) CyberIoT Chair of Excellence at the University of Brest for its support.

**Data Availability** Data sharing was not applicable to this article as no datasets were generated or analyzed during the current study.

## Declarations

**Conflict of interest** The authors have no conflicts of interest to declare that are relevant to the content of this article.

**Publisher's Note** Springer Nature remains neutral with regard to jurisdictional claims in published maps and institutional affiliations.